\documentclass[a4paper,11pt]{article}
\usepackage{jheppub} 
\usepackage{lineno}
\usepackage{braket}
\usepackage{longtable}

\newcommand{\FO}{\mathrm{FO}}
\newcommand{\phys}{\mathrm{phys}}
\newcommand{\LE}{\mathrm{LE}}
\newcommand{\univ}{\mathrm{univ}}
\newcommand{\PG}{\mathrm{PG}}
\newcommand{\PGT}{\mathrm{PGT}}
\newcommand{\Bel}{\mathrm{Bel}}
\newcommand{\can}{\mathrm{can}}

\title{Pseudo-Gauge Stabilizers and Fibration Structure of the Cooper--Frye Map at Freeze-Out}

\author{Jiahua Tian}
\affiliation{School of Physics, East China Normal University\\
500 Dongchuan Road, Shanghai, China}

\emailAdd{jtian1905@gmail.com}

\abstract{We study the pseudo-gauge transformation (PGT) freedom at freeze-out in relativistic spin hydrodynamics. The Cooper--Frye map is shown to factor through the quotient of freeze-out data by a universal stabilizer, yielding a stratified fibration over the space of thermodynamic Lagrange multipliers. This classifies observables into base and fiber types, bounds the number of independent PGT-sensitive observables by the family-restricted fiber dimension, and implies cross-observable consistency relations. Applied to heavy-ion polarization data, the fibration structure provides a structural interpretation of the tension between $\Lambda$ polarization and $\phi$-meson spin alignment as evidence that the vorticity-dominated response sector may need to be enlarged with local field-correlation data. We show that Weyl-anomaly-induced currents studied recently are classified as base observables and recover the known Belinfante--canonical obstruction $\Omega_{ab}\neq\varpi_{ab}$ from the stabilizer condition.}

\begin{document}
\maketitle
\flushbottom

\section{Introduction}\label{sec:intro}

The discovery of global $\Lambda$ hyperon polarization in non-central heavy-ion collisions~\cite{STAR:2017ckg, STAR:2018gyt} has established a quantitative bridge between the bulk rotational dynamics of the quark-gluon plasma (QGP) and the spin polarization of the produced hadrons. Subsequent observations of vector-meson spin alignment~\cite{STAR:2022fan} and of azimuthal-angle-dependent local spin polarization have stimulated an intense theoretical effort to understand spin transport in relativistic fluids and the freeze-out conversion of fluid spin data into hadronic observables~\cite{Becattini:2020ngo, Becattini:2022zvf}. Following the standard sharp-switching idealization of heavy-ion hydrodynamics, we model the spin-hydrodynamic stage of the QGP as a four-dimensional domain $M_4$ whose future boundary contains a three-dimensional freeze-out hypersurface $\Sigma_\FO$~\cite{Huovinen:2012is, Becattini:2020ngo}. On this hypersurface a Cooper--Frye prescription~\cite{Cooper:1974mv} converts surface hydrodynamic data into hadronic spectra and polarizations~\cite{Becattini:2013fla, Liu:2021nyg, Sheng:2025cjk, Huang:2024ffg}.

A persistent obstruction to making this conversion precise is the pseudo-gauge transformation (PGT) freedom of the spin tensor~\cite{Hehl:1976vr, Becattini:2011zlx, Becattini:2012pp, Becattini:2018duy, Speranza:2020ilk, Drogosz:2024rbd}. The decomposition of the conserved angular momentum current into an orbital and a spin part is not unique: different pseudo-gauges, related by a PGT generator $\Phi^{\lambda,\mu\nu}$, give different stress-energy and spin tensors that share the same total energy-momentum and angular momentum but differ at the level of local densities. Buzzegoli's explicit computation~\cite{Buzzegoli:2021wlg} establishes that the standard formulae for $\Lambda$ polarization and for the axial vortical conductivity are sensitive to this PGT choice. In particular, the canonical--Belinfante shift of the spin-polarization vector $S^\mu(k)$ contains the freeze-out integral of $(\varpi_{\mu\nu} - \Omega_{\mu\nu})$, where $\varpi_{\mu\nu}$ is the thermal vorticity and $\Omega_{\mu\nu}$ is the spin potential.

The novelty of the present work is not the observation that local-equilibrium spin observables can depend on the pseudo-gauge choice, but the formulation of this dependence as a universal stabilizer problem for the freeze-out map. In contrast to approaches that remove the ambiguity by imposing pseudo-gauge invariance directly on the local-equilibrium density operator, we keep the fixed-multiplier Cooper--Frye construction and identify the subgroup $H_\univ$ of pseudo-gauge transformations that is invisible to all admissible readouts. This leads to the quotient structure $\mathcal Q_\FO=R_\FO/H_\univ$, to a base/fiber classification of observables, and to a finite-family counting criterion for residual PGT-sensitive directions.

The key technical input is a Bogoliubov--Kubo--Mori (BKM) argument proving that the c-number shift condition on $\Delta_\Phi\hat K_\LE$ is both necessary and sufficient for observable invariance. The quotient $\mathcal{Q}_\FO$ is naturally fibered over the space $\mathcal{M}_\phys$ of thermodynamic Lagrange multipliers, and the framework predicts cross-observable consistency relations: for several fiber observables measured in the same kinematic bin, their simultaneous values must lie on the image of a common RSH and pseudo-gauge response slice. We apply this to the joint $\Lambda$-hyperon-polarization and $\phi$-meson-spin-alignment sector and show that the consistency condition implied by the fibration picture provides a structural explanation of the local field-correlation mechanism proposed in~\cite{Sheng:2022wsy}: the apparent tension with a narrow vorticity-dominated description is resolved by enlarging the freeze-out response space, with the added sector approximately invisible to $\Lambda$ polarization but visible to $\phi$-meson spin alignment. Finally, as an explicit check, the canonical-to-Belinfante PGT for a Dirac spin fluid on a planar $\Sigma_\FO$ recovers Buzzegoli's obstruction $\Omega_{ab}\neq\varpi_{ab}$ directly from the universal stabilizer condition.

The paper is organized as follows. In section~\ref{sec:setup} we set up the local-equilibrium chain, derive the central insertion formula~\eqref{eq:K_PGT}, and prove the universal stabilizer theorem. In section~\ref{sec:GE_full} we establish that the universal stabilizer saturates the full pseudo-gauge group at global thermodynamic equilibrium, and we carry out the canonical-to-Belinfante test as the simplest one-parameter linearization away from this saturated endpoint. Section~\ref{sec:fibration} develops the fibration picture and the resulting classification of physical observables. Section~\ref{sec:predictions} extracts the framework-level predictions. We conclude in section~\ref{sec:discussion}.

\section{Pseudo-Gauge Transformations and the Universal Stabilizer at Local Equilibrium}\label{sec:setup}

We consider quark-gluon plasma (QGP) described as a relativistic spin hydrodynamics (RSH) system in 4d spacetime $M_4$ with a sharp 3d freeze-out hypersurface $\Sigma_\FO = \partial M_4$~\cite{Huovinen:2012is, Becattini:2020ngo}. We then use the Cooper--Frye (CF) relation to map hydrodynamic variables on $\Sigma_\FO$ to genuine physical hadronic observables~\cite{Cooper:1974mv}. The RSH can be described by an action principle~\cite{Banerjee:2012iz, Jensen:2012jh, Gallegos:2021bzp, Gallegos:2022jow, Hongo:2021ona, Florkowski:2017ruc, Florkowski:2018fap}. More precisely, for a system in equilibrium on a stationary background with timelike Killing vector, we consider the equilibrium partition function $W$, a real local functional of the background sources organized in a derivative expansion~\cite{Jensen:2012jh}. The first variations of $W$ with respect to any source give the corresponding equilibrium currents through the integrated identity
\begin{equation}
    \delta W = \int d^4x\, \sqrt{-g}\,\bigl[\tfrac12 T^{\mu\nu}\delta g_{\mu\nu} + \tfrac12 S^{\mu,ab}\delta\omega_\mu{}^{ab} + j^\mu \delta A_\mu + \cdots\bigr] + \text{boundary terms}\,.
\end{equation}
A pseudo-gauge transformation (PGT) acts on the responses $T$, $S$, etc~\cite{Hehl:1976vr, Becattini:2018duy, Speranza:2020ilk}. More concretely, PGT acts on stress tensor $T$ and spin $S$ as:
\begin{equation}\label{eq:PGT}
    T^{\mu\nu} \rightarrow T^{\mu\nu} + \partial_\lambda X^{\lambda\mu\nu},\ S^{\lambda,\mu\nu} \rightarrow S^{\lambda,\mu\nu} - \Phi^{\lambda,\mu\nu}
\end{equation}
for $X^{\lambda\mu\nu} = \tfrac12(\Phi^{\lambda,\mu\nu} - \Phi^{\mu, \lambda\nu} - \Phi^{\nu, \lambda\mu})$ and $\Phi^{\lambda,\mu\nu} = -\Phi^{\lambda,\nu\mu}$. The current $j^\mu$ is PGT-invariant by construction. Thus the induced changes of the response operators on $\Sigma_{\FO}$, after contraction with the unit normal $n_\mu$, are:
\begin{equation}\label{eq:Delta_responses}
    \Delta_\Phi T_\Sigma^\nu = n_\mu \partial_\lambda X^{\lambda\mu\nu},\qquad \Delta_\Phi S_\Sigma^{\mu\nu} = -n_\lambda\Phi^{\lambda,\mu\nu},\qquad \Delta_\Phi j_\Sigma = 0.
\end{equation}

We collect the responses on $\Sigma_\FO$ into the abstract space $R_\FO$. For a chosen observable or measurement prescription $\mathcal O$, the physical question is how to construct a CF-type readout from $R_\FO$ to the corresponding value space $\mathcal M_\mathcal O$, which we denote
\begin{equation}\label{eq:CFmap}
    \mathfrak{F}_{\mathcal O}: R_\FO^{\LE} \;\longrightarrow\; \mathcal M_\mathcal O\,.
\end{equation}
For a fluid at local thermal equilibrium (LE)\footnote{The LE assumption is that the state of the fluid at freeze-out is the maximum von~Neumann entropy state subject to the constraints that the mean values of the conserved currents $\hat T^{\mu\nu}$, $\hat S^{\lambda,\mu\nu}$, $\hat j^\mu$ on $\Sigma_\FO$ match the physical macroscopic densities~\cite{Jaynes:1957zza, Jaynes:1957zz}; the Lagrange multipliers $(\beta^\mu,\Omega_{ab},\xi)$ are the conjugates of those constraints. Equivalently, the state is described by the density operator $\hat\rho_\LE = e^{-\hat K_\LE}/Z$ of~\eqref{eq:Define_rho} below, so that the entire freeze-out information content is encoded in $(\beta,\Omega,\xi)$ together with the operator content of $\hat K_\LE$. Non-equilibrium gradient corrections to $\hat\rho_\LE$, dissipative transport coefficients, and post-freeze-out hadronic rescattering are assumed to be subleading or to be treated as separate corrections~\cite{Zubarev:1979, Becattini:2014yxa}.}, $\mathfrak{F}_{\mathcal O}$ is constructed via the chain~\cite{Liu:2021nyg, Huang:2024ffg, Buzzegoli:2021wlg}:
\begin{equation}\label{eq:Chain}
    R_\FO^{\LE} \xrightarrow{\mathrm{PG\ choice}} \hat{\rho}_\LE \rightarrow W(x,p) \rightarrow \mathcal M_\mathcal O
\end{equation}
where $\hat{\rho}_\LE$ is the LE density operator on $\Sigma_\FO$, viewed as a functional of the Lagrange multipliers sourcing the surface responses~\cite{Buzzegoli:2021wlg, Becattini:2014yxa}, and $W(x,p)$ is the Wigner function constructed from $\hat{\rho}_\LE$~\cite{Huang:2024ffg}. A pseudo-gauge choice is implicitly fixed in the first arrow, since the operators $\hat T^{\mu\nu}$ and $\hat S^{\lambda,\mu\nu}$ entering $\hat\rho_\LE$ depend on the pseudo-gauge representative.

We denote the operator-valued representative of the freeze-out data in a given pseudo-gauge $\PG$ by
\begin{equation}
    \widehat{R}_{\FO}^{(\PG)} = \left( n_\mu \hat j^\mu,\, n_\mu \hat T_{(\PG)}^{\mu\nu},\, n_\mu \hat S_{(\PG)}^{\mu,ab},\, n_\mu \hat J_{(\PG)}^{\mu,ab} \right)
\end{equation}
where the parenthetical $\PG$ indicates pseudo-gauge dependence. Maximizing the von~Neumann entropy under the constraints of given mean densities of conserved currents on $\Sigma_\FO$~\cite{Zubarev:1979, Becattini:2014yxa, Huang:2024ffg} yields the local-equilibrium density operator:
\begin{equation}\label{eq:Define_rho}
    \hat\rho_{\LE}^{(\PG)} = \frac{1}{Z_{\LE}^{(\PG)}} \exp\!\left[-\hat K_{\LE}^{(\PG)}\right]
\end{equation}
with the local-equilibrium generator
\begin{equation}\label{eq:Define_K}
    \hat K_{\LE}^{(\PG)} = \int_{\Sigma_\FO} d\Sigma_\mu \left[ \hat T_{(\PG)}^{\mu\nu}\beta_\nu -\xi\,\hat j^\mu -\frac12 \Omega_{ab}\hat S_{(\PG)}^{\mu,ab} \right]\,.
\end{equation}
Here $\beta^\mu = u^\mu/T$ is the inverse-temperature four-vector, $\xi = \mu/T$ is the thermal chemical potential, and $\Omega_{ab}$ is the spin potential. These Lagrange multipliers are determined by the constraint that the LE mean values of $\hat T_{(\PG)}^{\mu\nu}$, $\hat S_{(\PG)}^{\mu,ab}$, $\hat j^\mu$ reproduce the physical surface densities~\cite{Becattini:2014yxa}. In the fibration setup below, we adopt the \emph{fixed-multiplier comparison scheme}: the thermodynamic Lagrange multipliers $(\beta,\xi,\Omega)$ are held fixed as externally prescribed fields on $\Sigma_\FO$ when comparing different pseudo-gauge representatives. This is a stated assumption, not a derived consequence: in a maximum-entropy construction the multipliers are conjugate to the constrained operators, and if those operators change under PGT the multipliers required to describe the same physical state could in principle change as well. The present paper studies the comparison at fixed $(\beta,\xi,\Omega)$; the pseudo-gauge dependence then resides entirely in which pair of operators is inserted into~\eqref{eq:Define_K}.

Throughout this paper we adopt the \emph{fundamental-$u$} convention: the fluid four-velocity $u^\mu$ (and hence $\beta^\mu=u^\mu/T$) is an independently specified thermodynamic variable, not derived from the stress tensor. In the alternative \emph{Landau-frame} convention, $u^\mu$ is defined as the timelike eigenvector of $T^{\mu\nu}$, so $\beta^\mu$ itself acquires a PGT-induced shift through $\Delta_\Phi T^{\mu\nu}=\partial_\lambda X^{\lambda\mu\nu}$. The universal stabilizer in the Landau reading would then carry the additional constraint $\partial_\lambda X^{\lambda\mu\nu}=0$ to ensure $\delta_\Phi\beta=0$, yielding a strictly smaller subgroup $H_\univ^{\rm L}\subseteq H_\univ$. Since the Landau frame is widely used in relativistic spin hydrodynamics phenomenology~\cite{Florkowski:2017ruc, Florkowski:2018fap, Buzzegoli:2021wlg}, all results of the present paper that depend on $H_\univ$ apply \emph{a fortiori} under the Landau-frame reading, where the stabilizer is smaller and the fiber is correspondingly larger.

\paragraph{Scope of the comparison.} The statements below should be read within this fixed-multiplier, fundamental-$u$ local-equilibrium scheme. The Cooper--Frye readout operators considered in the main text are PGT-invariant, so PGT acts on their expectation values through the density operator rather than through an explicit operator variation. The BKM non-degeneracy argument is used in a finite-volume/UV-regulated setting and then formally continued to the field-theory limit, as detailed in Appendix~\ref{app:conventions}. Finally, when we say that two observables are evaluated on the same physical state, we mean the same class $[R]\in\mathcal Q_\FO$, not merely the same thermodynamic multipliers $(\beta,\Omega,\xi)$.

We denote by
\begin{equation}\label{eq:LE_assignment}
    \mathcal L_\LE:\; R_\FO^{\LE}\longrightarrow \{\hat\rho_\LE\},
    \qquad
    R\longmapsto \hat\rho_\LE[R],
\end{equation}
the local-equilibrium assignment defined by~\eqref{eq:Define_rho}--\eqref{eq:Define_K}. Here $R_\FO^{\LE}\subset R_\FO$ is the sector of freeze-out data for which the local-equilibrium construction is valid: a point $R$ supplies, or determines by the entropy-maximization constraints, the thermal multipliers together with a pseudo-gauge representative of the surface response operators. This assignment is not the definition of $R_\FO$ itself, and it need not be injective on all freeze-out data. The quotient used in the fibration construction will therefore be taken directly on $R_\FO$, after identifying below which pseudo-gauge transformations leave the local-equilibrium density operator, and hence all observables, unchanged.

Applying~\eqref{eq:PGT} to $\hat K_\LE$ and holding the multipliers $(\beta,\xi,\Omega)$ fixed, we obtain the central insertion formula
\begin{equation}\label{eq:K_PGT}
    \Delta_\Phi \hat K_{\LE} = \int_{\Sigma_\FO} d\Sigma_\mu \left[ \beta_\nu\,\partial_\lambda \hat X^{\lambda\mu\nu} + \frac12\Omega_{ab}\hat\Phi^{\mu,ab} \right]\,.
\end{equation}
We have used that the current term $\xi\,\hat j^\mu$ is PGT-invariant. The first term is the divergence shift induced by~\eqref{eq:PGT} for the stress tensor, while the second is the direct shift of the spin sector. Equation~\eqref{eq:K_PGT} is the operator whose insertion into the BKM correlator with any observable controls all PGT-induced effects.

Consequently, the response of $\hat\rho_\LE$ to the perturbation~\eqref{eq:K_PGT} is~\cite{amari2000methods}
\begin{equation}\label{eq:Delta_rho}
    \Delta_\Phi \hat\rho_{\LE} = -\int_0^1 ds\, \hat\rho_{\LE}^{\,1-s} \left( \Delta_\Phi\hat K_{\LE} -\langle\Delta_\Phi\hat K_{\LE}\rangle_\LE \right) \hat\rho_{\LE}^{\,s}
\end{equation}
where $\hat\rho^s := \sum_n p_n^s \ket{n}\bra{n}$ when $\hat\rho = \sum_n p_n \ket{n}\bra{n}$. For any operator $\hat A$, we denote
\begin{equation}
    A := \langle \hat A\rangle_{\LE} = \mathrm{Tr}\!\left( \hat\rho_{\LE}\hat A \right)\,.
\end{equation}
Combining~\eqref{eq:Delta_rho} with~\eqref{eq:K_PGT}, and assuming that $\hat A$ is itself PGT-invariant, we obtain the static Kubo--Mori linear-response formula in the equivalent normalized-density-operator form~\cite{Kubo:1957mj, petz1993bogoliubov, Kubo:1991stat}:
\begin{equation}\label{eq:Change_of_A}
    \Delta_\Phi A \;=\; -\int_0^1 ds\, \mathrm{Tr}\!\left( \hat\rho_{\LE}^{\,1-s} \hat A \hat\rho_{\LE}^{\,s} \Delta_\Phi\hat K_\LE \right) + \langle \hat A\rangle_\LE \langle \Delta_\Phi\hat K_\LE \rangle_\LE \;=:\; - \langle \hat A ; \Delta_\Phi\hat K_\LE \rangle_\LE\,,
\end{equation}
where $(X, Y)_\rho := \int_0^1 ds\, \mathrm{Tr}\!\left(\rho^{1-s} X^\dagger \rho^{s} Y\right)$ is the Bogoliubov (or Kubo-Mori, BKM) inner product of $X$ and $Y$ with respect to the density operator $\rho$~\cite{amari2000methods, petz1993bogoliubov}. We refer to $\langle X; Y \rangle_\LE$ as the connected BKM correlator with respect to $\hat \rho_\LE$. Because the BKM form $\langle \cdot;\cdot\rangle_\LE$ is positive-definite on the space of operators modulo c-number multiples of the identity (see Appendix~\ref{app:conventions} for the proof in a finite-volume/UV-regulated setting, which we then use formally in the continuum limit), $\langle \hat A; \hat B\rangle_\LE$ vanishes for all $\hat A$ if and only if $\hat B \propto \mathbf 1$, a fact we will use repeatedly below.

Applying~\eqref{eq:Change_of_A} to the Wigner operator $\hat W(x,p)$ gives the PGT-induced shift of the Wigner function:
\begin{equation}\label{eq:Delta_W}
    \Delta_\Phi W(x,p) = -\langle \hat{W}(x,p);\, \Delta_\Phi \hat K_\LE \rangle_\LE\,.
\end{equation}
The Wigner function carries all phase-space information needed to compute spectra, polarization moments, and spin alignments at freeze-out~\cite{Liu:2020ymh, Sheng:2025cjk, Huang:2024ffg}.

The single-particle observables considered below can be written as suitable integrals of the Wigner function~\cite{Sheng:2025cjk}. We therefore consider a general functional
\begin{equation}\label{eq:Functional}
    \mathfrak{f}_{\mathcal O} : \mathcal{W} \;\longrightarrow\; \mathcal{M}_{\mathcal{O}}\,,
\end{equation}
where $\mathcal{W}$ is the space of admissible Wigner functions on $\Sigma_\FO$ and $\mathcal{M}_\mathcal{O}$ the space of physical observable values. The Wigner-space Cooper--Frye functional $\mathfrak f_{\mathcal O}$ depends on the specific observable; e.g., for hadron spectra it reduces to the standard $\int d\Sigma_\mu p^\mu f(x,p)$ integral, while for $\Lambda$ polarization it is the spin-axial projection of~\cite{Liu:2021nyg, Buzzegoli:2021wlg}.\footnote{Concretely, the observable-dependent readout may be the invariant spectrum $E_p dN/d^3p$ obtained from the Cooper--Frye flux integral~\cite{Cooper:1974mv,Huovinen:2012is}, the mean spin vector $S^\mu(p)$ obtained by projecting the Wigner function onto the axial/spin sector~\cite{Liu:2021nyg,Buzzegoli:2021wlg}, the vector-meson spin-alignment matrix element $\rho_{00}$ extracted from the decay angular distribution~\cite{Schilling:1969um,STAR:2022fan}, or the axial current $j_5^\mu$ and associated axial-vortical response obtained from the axial Wigner projection~\cite{Buzzegoli:2021wlg}. These examples differ only in the readout kernel and normalization applied to the same underlying phase-space data.} For small variations of $W$,
\begin{equation}
    \mathfrak{f}_{\mathcal O}[W + \epsilon \delta W] = \mathfrak{f}_{\mathcal O}[W] + \epsilon\, D\mathfrak{f}_{\mathcal O}|_W[\delta W]
\end{equation}
where $D\mathfrak{f}_{\mathcal O}|_W$ is the functional derivative of $\mathfrak{f}_{\mathcal O}$ at $W$. Combining with~\eqref{eq:Delta_W}:
\begin{equation}\label{eq:Delta_Ophys}
    \Delta_\Phi\bigl(\mathfrak{f}_{\mathcal O}[W]\bigr)
    =
    D\mathfrak{f}_{\mathcal O}|_W[\Delta_\Phi W]
    =
    -D\mathfrak{f}_{\mathcal O}|_W \!\left[\langle \hat W (x,p);\, \Delta_\Phi \hat K_\LE \rangle_\LE\right]\,.
\end{equation}
For a chosen observable, the local-equilibrium construction factors through the Wigner representation as
\begin{equation}\label{eq:CF_composite}
    R_\FO^{\LE}
    \xrightarrow{\;\mathcal L_\LE\;}
    \{\hat\rho_\LE\}
    \xrightarrow{\;\mathrm{Wig}\;}
    \mathcal W
    \xrightarrow{\;\mathfrak f_{\mathcal O}\;}
    \mathcal M_\mathcal O\,,
\end{equation}
where $\mathrm{Wig}$ denotes the construction of the Wigner function from $\hat\rho_\LE$. Thus the composite readout in~\eqref{eq:CFmap} is $\mathfrak F_{\mathcal O}=\mathfrak f_{\mathcal O}\circ\mathrm{Wig}\circ\mathcal L_\LE$. This equation only isolates the intermediate phase-space representation used to evaluate observables: the response operator $\chi_W$ and the observable derivative $D\mathfrak f_{\mathcal O}|_W$ both act at this Wigner-function stage.

This is the master formula relating the PGT generator $\Phi$ to the change of any physical observable value. The structure is striking: the entire PGT dependence sits in the BKM correlator $\langle \hat W;\,\Delta_\Phi\hat K_\LE\rangle_\LE$, and the dependence on the choice of observable enters only through the linear functional $D\mathfrak{f}_{\mathcal O}|_W$ at the very end~\footnote{This statement assumes that the measured operator or Cooper--Frye readout functional is PGT-invariant, so that PGT acts only through the density operator. This is the case for the particle and current observables used below, such as $\Lambda$ polarization, $\rho_{00}^{\phi}$, the axial current, and the Weyl-anomaly vector current. For an explicitly transforming local representative, such as a local stress tensor, spin tensor, or separated orbital/spin angular-momentum density, one must add the direct operator variation at fixed density operator to the state-induced response $-\langle\hat A;\Delta_\Phi\hat K_\LE\rangle_\LE$.}. The notation $\chi_W(X) := -\langle \hat W; X\rangle_\LE$ identifies the BKM correlator with a linear map $\chi_W:\mathcal{A}\to\mathcal{W}$ from the operator algebra to Wigner-function fluctuations, and~\eqref{eq:Delta_Ophys} reads $\Delta_\Phi(\mathfrak f_{\mathcal O}[W]) = D\mathfrak{f}_{\mathcal O}|_W\circ\chi_W(\Delta_\Phi\hat K_\LE)$. The stabilizer of an observable $\mathcal{O}$ is therefore
\begin{equation}\label{eq:HO_stabilizer}
    H_{\mathcal{O}} := \{ \Phi \,|\, \Delta_\Phi\hat K_\LE \in \ker(D\mathfrak{f}_{\mathcal O}|_W \circ \chi_W) \}\,.
\end{equation}
This is observable-dependent through both factors $D\mathfrak{f}_{\mathcal O}|_W$ and $\chi_W$.

Since the Wigner-space functional $\mathfrak{f}_{\mathcal O}$ depends on $\mathcal{O}$, $H_{\mathcal{O}}$ varies even for fixed $W$. It is therefore natural to ask whether there is a {\em universal} subgroup $H_\univ$ contained in every $H_\mathcal{O}$ and which depends only on the LE density operator, not on any specific observable.\label{sec:Huniv} Indeed, from the definition~\eqref{eq:Define_rho}, the shift
\begin{equation}\label{eq:Shift_c}
    \hat K_\LE \rightarrow \hat K_\LE + c_\Phi \mathbf{1}
\end{equation}
multiplies $e^{-\hat K_\LE}$ by the scalar $e^{-c_\Phi}$, which cancels exactly between numerator and denominator of $\hat\rho_\LE$. Therefore the entire density operator is unchanged, and so is every observable $\langle\hat A\rangle_\LE = \mathrm{Tr}(\hat\rho_\LE\hat A)$. We are thus led to define
\begin{equation}\label{eq:Univ_Stab}
    H_\univ = \{ \Phi \,|\, \Delta_\Phi \hat K_\LE = c_\Phi \mathbf{1} \}\,.
\end{equation}
Using~\eqref{eq:K_PGT}, this is the explicit subgroup
\begin{equation}\label{eq:Def_Huniv}
    H_\univ = \left\{\,\Phi \,\Big|\, \int_{\Sigma_\FO} d\Sigma_\mu \left[ \beta_\nu\,\partial_\lambda \hat X^{\lambda\mu\nu} + \frac12\Omega_{ab}\hat\Phi^{\mu,ab} \right] = c_\Phi\mathbf 1\,\right\}\,.
\end{equation}

The condition~\eqref{eq:Univ_Stab} is in fact both necessary and sufficient for an observable to be PGT-invariant for every choice of $\hat A$. From~\eqref{eq:Change_of_A},
\begin{equation}
    \Delta_\Phi\langle\hat A\rangle_\LE = 0\;\text{for all hermitian}\;\hat A \;\Longleftrightarrow\; \langle \hat A;\, \Delta_\Phi\hat K_\LE\rangle_\LE = 0\;\text{for all}\;\hat A\,.
\end{equation}
Since the BKM inner product is non-degenerate on the operator algebra modulo c-number multiples of the identity~\cite{petz1993bogoliubov, petz2007quantum}, the right-hand side holds if and only if $\Delta_\Phi\hat K_\LE \in \mathbb{C}\,\mathbf 1$. Hence
\begin{equation}\label{eq:Huniv_iff}
    \Phi \in H_\univ \;\Longleftrightarrow\; \Delta_\Phi\langle\hat A\rangle_\LE = 0 \;\text{for every hermitian}\;\hat A\,.
\end{equation}
$H_\univ$ is therefore the exact universal stabilizer for the full expectation-value algebra, not merely a sufficient subgroup. For restricted observable families (e.g., only spin polarization moments), the actual common stabilizer $\bigcap_{\mathcal{O}}H_\mathcal{O}$ can be larger than $H_\univ$, in which case~\eqref{eq:Univ_Stab} provides a guaranteed lower bound.

Since PGTs in $H_\univ$ do not affect any physical observable, the CF map $\mathfrak{F}_{\mathcal O}$ is constant on $H_\univ$-orbits in $R_\FO$ for every chosen observable $\mathcal O$ and therefore factorizes through the quotient. There exists a unique induced map
\begin{equation}\label{eq:Factorization}
    \widetilde{\mathfrak{F}}_{\mathcal O}\,:\;\mathcal{Q}_\FO := R_\FO/H_\univ \;\longrightarrow\; \mathcal{M}_\mathcal{O}\,, \qquad \mathfrak{F}_{\mathcal O} = \widetilde{\mathfrak{F}}_{\mathcal O}\circ q\,,
\end{equation}
where $q:R_\FO\twoheadrightarrow\mathcal{Q}_\FO$ is the quotient map.

The Lagrange multipliers $(\beta,\xi,\Omega)$ enter the condition~\eqref{eq:Def_Huniv} explicitly through the coefficients $\beta_\nu$ and $\Omega_{ab}$. Consequently, the universal stabilizer is state-dependent: at different points $b = (\beta,\xi,\Omega)\in\mathcal{M}_\phys$, the subgroup $H_\univ[b]\subset G_\PGT$ takes different forms. At one extreme, when $(\beta,\Omega)$ correspond to global equilibrium, namely $\beta^\mu$ a Killing vector field $\beta^\mu(x) = b^\mu + \varpi^\mu{}_\nu x^\nu$ with constant antisymmetric $\varpi_{\mu\nu}$~\cite{Becattini:2014yxa} and spin potential equal to the thermal vorticity, $\Omega_{ab}=\varpi_{ab}$, the entire group $G_\PGT$ lies in $H_\univ[b]$; this is shown explicitly in the next section. At the other extreme, for a generic local equilibrium state with $\Omega\neq\varpi$, $H_\univ[b]$ shrinks to a strictly proper subgroup of $G_\PGT$. This state dependence is the geometric content of the dependence of observables on the choices of pseudo-gauge exemplified in~\cite{Buzzegoli:2021wlg} and will be developed further in section~\ref{sec:fibration}.

\section{Global Equilibrium and the Belinfante--Canonical Test}\label{sec:GE_full}

This section carries out the basic tests of the universal stabilizer construction. We first show, in section~\ref{sec:GTE_saturation}, that at global thermodynamic equilibrium $H_\univ$ saturates the full pseudo-gauge group, so that pseudo-gauge ambiguity has no physical consequence in this limit. We then linearize away from this saturated endpoint along the simplest one-parameter family of pseudo-gauges, the Belinfante--canonical line for a Dirac spin fluid~\footnote{The word ``line'' is our shorthand for the one-parameter interpolation in pseudo-gauge space obtained by applying the PGT generator $\Phi_\tau^{\lambda,\mu\nu}=\tau S_\can^{\lambda,\mu\nu}$ to the canonical Dirac pair, with $0\leq\tau\leq1$. The endpoint $\tau=0$ is the canonical pseudo-gauge, while $\tau=1$ is the Belinfante pseudo-gauge, for which the spin tensor is absorbed into the Belinfante-Rosenfeld symmetric stress tensor. The endpoints and the corresponding pseudo-gauge transformation are standard~\cite{Belinfante:1940,Rosenfeld:1940,Hehl:1976vr,Speranza:2020ilk,Buzzegoli:2021wlg}, and the phrase ``Belinfante--canonical line'' is a naming convention used here.}, and identify the precise obstruction to $H_\univ$ membership that emerges away from equilibrium. The two parts share a common logical structure: at the equilibrium endpoint the spin potential equals the thermal vorticity and the entire $G_\PGT$ acts trivially; away from this endpoint the residue $\Omega - \varpi$ controls the failure of $H_\univ$ to absorb the Belinfante--canonical PGT. The predictions of section~\ref{sec:predictions} rest on the explicit computations of this section.

\subsection{Saturation at global equilibrium}\label{sec:GTE_saturation}

In this subsection we derive the claim that $H_\univ = G_\PGT$ at global equilibrium mentioned in section~\ref{sec:Huniv}. The argument proceeds by showing that at global equilibrium the generator $\hat K_\LE$ reduces to a linear combination of the total conserved charges of the Poincar\'e and baryon-number symmetries, and that these total charges are exactly pseudo-gauge invariant on $\Sigma_\FO\cong\mathbb{R}^3$ with standard fall-off, so that $\Delta_\Phi\hat K_\LE^{\rm GE}=0$ and every $\Phi$ lies in $H_\univ$.

Global thermodynamic equilibrium for a relativistic fluid with inverse temperature four-vector $\beta^\mu = u^\mu/T$ is characterized by the Killing condition~\cite{Becattini:2014yxa}
\begin{equation}\label{eq:Killing}
    \partial_{(\mu}\beta_{\nu)} = 0\,,
\end{equation}
which in flat spacetime is solved by
\begin{equation}\label{eq:beta_general}
    \beta^\mu(x) = b^\mu + \varpi^\mu{}_\nu\,x^\nu\,, \qquad \varpi_{\mu\nu} = -\partial_{[\mu}\beta_{\nu]} = \text{constant antisymmetric}\,.
\end{equation}
This describes a uniformly rotating thermal fluid; the static case is the special instance $\varpi_{\mu\nu} = 0$. Stationarity of $\hat\rho_\LE$ under the dynamics then forces the spin potential at global equilibrium to be equal to the thermal vorticity~\cite{Becattini:2018duy, Becattini:2022zvf, Buzzegoli:2021wlg, Becattini:2025oyi},
\begin{equation}\label{eq:Locking_GE}
    \Omega_{ab} = \varpi_{ab}\,.
\end{equation}
The reason is that orbital and spin angular momenta are not separately conserved in an interacting theory; only their sum $\hat J^{ab} = \hat L^{ab} + \hat S^{ab}$ commutes with the Hamiltonian. The combination $-\tfrac{1}{2}\varpi_{ab}\hat L^{ab} - \tfrac{1}{2}\Omega_{ab}\hat S^{ab}$ that would otherwise appear in $\hat K_\LE$ is therefore a symmetry generator only when the coefficients in front of $\hat L^{ab}$ and $\hat S^{ab}$ coincide, i.e. when $\Omega_{ab}=\varpi_{ab}$. We will make this argument more explicit in a moment.

Inserting~\eqref{eq:beta_general} and~\eqref{eq:Locking_GE} into the explicit definition~\eqref{eq:Define_K} of $\hat K_\LE$, and using the constancy of the Lagrange multipliers $b^\mu$, $\varpi_{ab}$, $\xi$ at global equilibrium, the generator takes the form
\begin{equation}\label{eq:K_GE_split}
    \hat K_\LE \,=\, \int_\Sigma d\Sigma_\mu\,\hat T^{\mu\nu}\bigl(b_\nu + \varpi_{\nu\rho}x^\rho\bigr) \,-\, \xi\!\int_\Sigma d\Sigma_\mu\,\hat j^\mu \,-\, \tfrac{1}{2}\varpi_{ab}\!\int_\Sigma d\Sigma_\mu\,\hat S^{\mu,ab}\,.
\end{equation}
The $b_\nu$ piece of the stress-tensor term gives the total energy-momentum directly. The $x$-dependent piece is rearranged using the antisymmetry of $\varpi_{\nu\rho}$,
\begin{equation}\label{eq:xT_rearrangement}
    \int_\Sigma d\Sigma_\mu\,\hat T^{\mu\nu}\,\varpi_{\nu\rho}\,x^\rho \,=\, \tfrac{1}{2}\,\varpi_{\nu\rho}\!\int_\Sigma d\Sigma_\mu\bigl(x^\rho\hat T^{\mu\nu} - x^\nu\hat T^{\mu\rho}\bigr) \,=\, -\tfrac{1}{2}\,\varpi_{\nu\rho}\,\hat L^{\nu\rho}\,,
\end{equation}
which identifies the integrated orbital angular momentum. Combining~\eqref{eq:xT_rearrangement} with the spin-tensor piece of~\eqref{eq:K_GE_split} via the total angular momentum $\hat J^{\nu\rho} = \hat L^{\nu\rho} + \hat S^{\nu\rho}$, one obtains the compact form
\begin{equation}\label{eq:K_GE_compact}
    \hat K_\LE^{\rm GE} \;=\; b_\nu\,\hat P^\nu \,-\, \xi\,\hat Q \,-\, \tfrac{1}{2}\,\varpi_{\nu\rho}\,\hat J^{\nu\rho}\,,
\end{equation}
where the integrated total charges are defined as
\begin{equation}\label{eq:total_charges}
    \hat P^\nu \,=\, \int_\Sigma d\Sigma_\mu\,\hat T^{\mu\nu}\,,\qquad \hat J^{\nu\rho} \,=\, \int_\Sigma d\Sigma_\mu\bigl(x^\nu\hat T^{\mu\rho} - x^\rho\hat T^{\mu\nu} + \hat S^{\mu,\nu\rho}\bigr)\,,\qquad \hat Q \,=\, \int_\Sigma d\Sigma_\mu\,\hat j^\mu\,.
\end{equation}
Equation~\eqref{eq:K_GE_compact} is the standard covariant form of the global-equilibrium statistical operator studied in~\cite{Becattini:2014yxa, Becattini:2020ngo, Zubarev:1979}, where the generator is built entirely from the total charges of the Poincar\'e algebra together with the baryon-number generator.

It remains to establish that each of these total charges is pseudo-gauge invariant on $\Sigma_\FO \cong \mathbb{R}^3$ with the standard fall-off conditions on the fields at spatial infinity. The baryon charge $\hat Q$ is invariant by construction, since $\hat j^\mu$ does not transform under~\eqref{eq:PGT}. For the total momentum, the shift induced by the PGT is a 4-divergence $\partial_\lambda\hat X^{\lambda 0\nu}$ integrated against $d^3x$; antisymmetry of $\hat X^{\lambda\mu\nu}$ in its first pair kills the $\partial_0\hat X^{00\nu}$ piece, and the spatial divergence $\partial_i \hat X^{i0\nu}$ reduces by Gauss's theorem to a surface integral on $S^2_\infty$ that vanishes whenever $\hat\Phi^{\lambda,\mu\nu}$ falls off faster than $1/r^2$, the standard assumption on pseudo-gauge generators acting on states of finite total energy-momentum. Hence
\begin{equation}\label{eq:DeltaP_zero}
    \Delta_\Phi \hat P^\nu \,=\, 0\,.
\end{equation}

The total angular momentum requires more care. The shift of its local density is
\begin{equation}\label{eq:DeltaJdensity}
    \Delta_\Phi\!\bigl(x^\mu \hat T^{\lambda\nu} - x^\nu \hat T^{\lambda\mu} + \hat S^{\lambda,\mu\nu}\bigr) \,=\, \partial_\rho\bigl[x^\mu \hat X^{\rho,\lambda\nu} - x^\nu \hat X^{\rho,\lambda\mu}\bigr] \,+\, \mathcal{R}^{\lambda,\mu\nu}\,,
\end{equation}
where the first term is a total derivative obtained by the Leibniz identity $\partial_\rho(x^\mu A^\rho) = A^\mu + x^\mu\partial_\rho A^\rho$ and the residual is
\begin{equation}
    \mathcal{R}^{\lambda,\mu\nu} \,=\, -\hat X^{\mu,\lambda\nu} \,+\, \hat X^{\nu,\lambda\mu} \,-\, \hat\Phi^{\lambda,\mu\nu}\,.
\end{equation}
Substituting the definition $\hat X^{\lambda\mu\nu} = \tfrac12(\hat\Phi^{\lambda,\mu\nu}-\hat\Phi^{\mu,\lambda\nu}-\hat\Phi^{\nu,\lambda\mu})$ and using the antisymmetry $\hat\Phi^{\alpha,\beta\gamma} = -\hat\Phi^{\alpha,\gamma\beta}$ in the last pair, one finds after a few lines
\begin{equation}\label{eq:R_evaluation}
    -\hat X^{\mu,\lambda\nu} + \hat X^{\nu,\lambda\mu} \,=\, \hat\Phi^{\lambda,\mu\nu}\,,
\end{equation}
so that the residual vanishes identically, $\mathcal{R}^{\lambda,\mu\nu} = 0$. The shift of the angular momentum density is therefore a pure total derivative. Integrating~\eqref{eq:DeltaJdensity} over $\Sigma_\FO$, the total derivative becomes a boundary integral on $S^2_\infty$ that vanishes by the same fall-off argument as for $\hat P^\nu$. Hence
\begin{equation}\label{eq:DeltaJ_zero}
    \Delta_\Phi \hat J^{\mu\nu} \,=\, 0\,.
\end{equation}

Combining~\eqref{eq:DeltaP_zero} and~\eqref{eq:DeltaJ_zero} in~\eqref{eq:K_GE_compact}, the PGT-induced shift of the global-equilibrium generator is
\begin{equation}\label{eq:DeltaK_GE}
    \Delta_\Phi \hat K_\LE^{\rm GE} \,=\, b_\nu\cdot 0 - \xi\cdot 0 - \tfrac{1}{2}\,\varpi_{\nu\rho}\cdot 0 \,=\, 0\,,
\end{equation}
which vanishes identically (and in particular equals $0\cdot\mathbf{1}$) for every $\Phi\in G_\PGT$. By the definition~\eqref{eq:Univ_Stab} we conclude
\begin{equation}\label{eq:Huniv_GE}
    H_\univ\bigl[\beta_{\rm Killing},\,\Omega = \varpi\bigr] \,=\, G_\PGT\,.
\end{equation}
We summarize the result as the following.

\noindent{\bf Proposition.} {\em Let the freeze-out hypersurface $\Sigma_\FO$ be diffeomorphic to $\mathbb{R}^3$ and assume that the pseudo-gauge generators $\hat\Phi^{\lambda,\mu\nu}$ decay faster than $1/r^2$ at spatial infinity. At a global-equilibrium state characterized by $\beta^\mu(x) = b^\mu + \varpi^\mu{}_\nu x^\nu$ with constant antisymmetric $\varpi_{\mu\nu}$, $\Omega_{ab} = \varpi_{ab}$, and constant chemical potential $\xi$, the universal stabilizer saturates the full pseudo-gauge group, $H_\univ = G_\PGT$. Equivalently, every Cooper--Frye observable map $\widetilde{\mathfrak F}_{\mathcal O}$ descends in this limit all the way to the thermodynamic moduli space $\mathcal{M}_\phys$, the fiber $G_\PGT/H_\univ$ degenerates to a point, and pseudo-gauge ambiguity has no physical consequence.}

The proposition makes precise the long-standing observation that the total Poincar\'e charges are PGT-invariant up to boundary terms~\cite{Hehl:1976vr, Becattini:2020ngo, Becattini:2018duy}. The equilibrium relation~\eqref{eq:Locking_GE} is essential: when $\Omega \neq \varpi$, the spin sector no longer combines with the orbital sector into the full $\hat J^{ab}$, the stabilizer strictly shrinks, and the canonical-to-Belinfante test below provides a simple one-parameter manifestation of this shrinking.

\subsection{The Belinfante--canonical test}\label{sec:BC_test}

For a Dirac field, the canonical spin tensor is totally antisymmetric in its three indices, $\hat S^{\lambda,\mu\nu}_\can = -\hat S^{\mu,\lambda\nu}_\can = -\hat S^{\lambda,\nu\mu}_\can$. The Belinfante symmetrization procedure~\cite{Belinfante:1940, Rosenfeld:1940, Hehl:1976vr} replaces the canonical pair $(\hat T_\can, \hat S_\can)$ by the symmetric Belinfante pair $(\hat T_\Bel, 0)$ and corresponds to the PGT generator
\begin{equation}\label{eq:Phi_Bel}
    \hat\Phi^{\lambda,\mu\nu} \;=\; \hat S_\can^{\lambda,\mu\nu}\,, \qquad \hat S^{\lambda,\mu\nu}_\Bel = 0\,.
\end{equation}
The Belinfante-symmetrized tensor combination is
\begin{equation}
    \hat X^{\lambda\mu\nu} \;=\; \tfrac12\bigl(\hat S_\can^{\lambda,\mu\nu} - \hat S_\can^{\mu,\lambda\nu} - \hat S_\can^{\nu,\lambda\mu}\bigr) \;=\; \tfrac12\,\hat S_\can^{\lambda,\mu\nu}\,,
\end{equation}
where the second equality uses total antisymmetry of $\hat S_\can$, so that the second and third terms cancel against each other. Substituting~\eqref{eq:Phi_Bel} into the master formula~\eqref{eq:K_PGT}, the canonical-to-Belinfante insertion is
\begin{equation}\label{eq:Delta_K_BC}
    \Delta_\Phi \hat K_\LE \;=\; \int_{\Sigma_\FO} d\Sigma_\mu \!\left[\beta_\nu \,\partial_\lambda \hat X^{\lambda\mu\nu} + \tfrac12 \Omega_{ab}\hat S_\can^{\mu,ab}\right]\,.
\end{equation}
The first piece can be integrated by parts. On a planar $\Sigma_\FO\cong\mathbb{R}^3$, the integration by parts produces
\begin{equation}
    \int_{\Sigma_\FO} d\Sigma_\mu\,\beta_\nu\,\partial_\lambda\hat X^{\lambda\mu\nu} \;=\; -\int_{\Sigma_\FO} d\Sigma_\mu\,(\partial_\lambda\beta_\nu)\,\hat X^{\lambda\mu\nu}
\end{equation}
modulo edge terms that vanish for fields with appropriate fall-off at spatial infinity. Decomposing $\partial_\lambda\beta_\nu = \xi_{\lambda\nu}-\varpi_{\lambda\nu}$ into the symmetric thermal shear\footnote{The thermal shear $\xi_{\lambda\nu}:=\partial_{(\lambda}\beta_{\nu)}$ is the symmetric part of the gradient of the inverse-temperature four-vector $\beta^\mu=u^\mu/T$, complementary to the antisymmetric thermal vorticity $\varpi_{\lambda\nu}=-\partial_{[\lambda}\beta_{\nu]}$. It is a standard ingredient of relativistic spin hydrodynamics at local equilibrium~\cite{Becattini:2021suc, Becattini:2022zvf} and vanishes identically at global equilibrium by the Killing condition. In the canonical-to-Belinfante test of this section, the thermal-shear piece arising from the integration by parts contracts against $\hat S_\can^{\lambda,\mu\nu}$, which is antisymmetric in $(\lambda,\nu)$; the contraction with the symmetric $\xi_{\lambda\nu}$ therefore vanishes identically, and only the antisymmetric (thermal-vorticity) part survives in~\eqref{eq:Delta_K_BC_simplified}. Beyond the canonical-to-Belinfante test, the thermal shear enters spin polarization predictions via the Wigner-function side of the BKM correlator and produces the spin-thermal-shear coupling identified in~\cite{Becattini:2021suc}.} and antisymmetric thermal vorticity parts, and using the total antisymmetry of $\hat S_\can^{\lambda,\mu\nu}$, only the antisymmetric component contracts non-trivially, giving
\begin{equation}\label{eq:Delta_K_BC_simplified}
    \Delta_\Phi \hat K_\LE \;=\; \tfrac12\int_{\Sigma_\FO} d\Sigma_\mu\,\bigl(\Omega_{ab}-\varpi_{ab}\bigr)\hat S_\can^{\mu,ab} + \text{(edge piece)}\,.
\end{equation}
For a planar $\Sigma_\FO$ with no edge contribution, the canonical-to-Belinfante insertion reduces to the compact expression
\begin{equation}\label{eq:KeyDelta}
    \Delta_\Phi^{\rm c\to B} \hat K_\LE \;=\; \tfrac12 \int_{\Sigma_\FO} d\Sigma_\mu\,(\Omega_{ab}-\varpi_{ab})\,\hat S_\can^{\mu,ab}\,.
\end{equation}

From~\eqref{eq:KeyDelta}, the canonical-to-Belinfante PGT belongs to $H_\univ$ if and only if the right-hand side equals $c\,\mathbf 1$ as an operator. Since $\hat S_\can^{\mu,ab}$ is not proportional to the identity in general, this requires the pointwise vanishing of the coefficient on the support of $\Sigma_\FO$:
\begin{equation}\label{eq:Locking}
    \Omega_{ab}(x) \;=\; \varpi_{ab}(x)\qquad \text{for all}\; x\in\Sigma_\FO\,.
\end{equation}
This is precisely the equilibrium relation between spin potential and thermal vorticity that we encountered in section~\ref{sec:GTE_saturation}, where $\Omega_{ab}=\varpi_{ab}$ was derived from the stationarity of $\hat\rho_\LE$~\cite{Becattini:2014yxa, Buzzegoli:2021wlg}. We have therefore shown:

\noindent{\bf Proposition.} {\em For a Dirac spin fluid on a planar freeze-out surface, ignoring edge contributions, the canonical-to-Belinfante PGT $\Phi^{\lambda,\mu\nu}=\hat S_\can^{\lambda,\mu\nu}$ belongs to the universal stabilizer $H_\univ$ if and only if $\Omega_{ab}=\varpi_{ab}$ on $\Sigma_\FO$.}

Conversely, if $\Omega_{ab}\neq\varpi_{ab}$ on a finite portion of $\Sigma_\FO$, the right-hand side of~\eqref{eq:KeyDelta} is a non-trivial integrated operator that is generically not proportional to the identity, hence $\Phi_\Bel\notin H_\univ$. By the master formula~\eqref{eq:Delta_Ophys}, all spin observables that probe the BKM correlator $\langle\hat W;\Delta_\Phi\hat K_\LE\rangle_\LE$ with non-vanishing weight in the spin direction will inherit a pseudo-gauge-dependent shift. The explicit form of the polarization shift, proportional to $\int_\Sigma d\Sigma\!\cdot\! p\,n_F(1-n_F)(\varpi_{ab}-\Omega_{ab})$ contracted with kinematic factors, is precisely the canonical-to-Belinfante difference computed in~\cite{Buzzegoli:2021wlg}. The universal-stabilizer framework therefore reproduces and geometrizes Buzzegoli's result.

\section{Fibration Structure of Equivalence Relations}\label{sec:fibration}

Two spaces are involved in the chain~\eqref{eq:Chain} and must be carefully distinguished. The first is the space of thermodynamic Lagrange multipliers:
\begin{equation}\label{eq:Mphys}
    \mathcal{M}_{\phys} \;:=\; \bigl\{\, (\beta^\mu(x),\,\Omega_{ab}(x),\,\xi(x))\;\text{on}\;\Sigma_\FO \,\bigr\}\,.
\end{equation}
These parameterize the local-equilibrium state and are held fixed across pseudo-gauge representatives in the comparison scheme adopted here~\cite{Becattini:2014yxa}. The second is the universal pseudo-gauge quotient of freeze-out data,
\begin{equation}\label{eq:Inequiv_rho}
    \mathcal{Q}_\FO \;:=\; R_\FO/H_\univ\,,
\end{equation}
which collapses pseudo-gauge representatives related by the universal stabilizer~\eqref{eq:Univ_Stab}. At fixed thermodynamic state $(\beta,\Omega,\xi)$, different pseudo-gauge choices generically give different density operators, and hence different physical observables~\cite{Buzzegoli:2021wlg}: this is precisely the statement that $\mathcal{M}_\phys$ and $\mathcal{Q}_\FO$ are not the same object. Note that $\mathcal{Q}_\FO=R_\FO/H_\univ$ removes only pseudo-gauge redundancies. It does not quotient by other possible parametrization redundancies of the freeze-out description, such as reparametrizations of $\Sigma_\FO$, local tetrad-frame choices, hydrodynamic field redefinitions, or operator-basis identities that leave $\hat K_\LE$ unchanged up to a c-number. Thus $\mathcal{Q}_\FO$ is the universal PGT quotient on which admissible Cooper--Frye observables are well-defined, but it is not asserted to be the minimal space of physically distinct local-equilibrium density operators.

It is convenient to fix terminology here, as it will be used throughout the rest of the paper. By a {\em physical state} on $\Sigma_\FO$ we mean the full local-equilibrium specification needed to determine $\hat\rho_\LE$, i.e., a point $[R]\in\mathcal{Q}_\FO$ rather than only a point of $\mathcal{M}_\phys$. The thermal Lagrange multipliers $(\beta,\Omega,\xi)$ alone do not fix $\hat\rho_\LE$: the local-equilibrium construction~\eqref{eq:Define_rho} additionally requires committing to a pseudo-gauge representative for the response operators that enter $\hat K_\LE$. The non-triviality of the fibration structure developed below is precisely the statement that this pseudo-gauge information is needed {\em beyond} the thermodynamic data to fix observable values; in the language of the projection $\pi$ defined just below, $\pi([R])\in\mathcal{M}_\phys$ recovers the thermal component of the physical state while the fiber direction records its residual pseudo-gauge component.

There is a natural projection
\begin{equation}\label{eq:Projection_pi}
    \pi:\; \mathcal{Q}_\FO \;\twoheadrightarrow\; \mathcal{M}_{\phys}
\end{equation}
obtained by forgetting the residual pseudo-gauge class and remembering only the Lagrange multipliers. In the pseudo-gauge orbit sector of $R_\FO$ considered here, $\pi$ is a quotient projection with fiber
\begin{equation}\label{eq:Fiber}
    F_{(\beta,\Omega,\xi)} \;=\; G_\PGT\big/H_\univ[\beta,\Omega,\xi]\,,
\end{equation}
the coset space of pseudo-gauge transformations modulo those that leave the density operator invariant at the given thermodynamic state. Equivalently, for each set of thermal parameters in $\mathcal{M}_\phys$, this fiber enumerates the physically inequivalent pseudo-gauge representatives in $\mathcal{Q}_\FO$ over that point, since within a single $H_\univ$-orbit all representatives produce identical $\hat\rho_\LE$ via~\eqref{eq:Inequiv_rho}. Because $H_\univ$ depends on $(\beta,\Omega,\xi)$ through~\eqref{eq:Def_Huniv}, the fibration is non-trivial, with the fiber varying with the base point. The pseudo-gauge part of the fibration is
\begin{equation}\label{eq:Fibration}
    G_\PGT/H_\univ[\beta,\Omega,\xi] \;\hookrightarrow\; \mathcal{Q}_\FO \;\xrightarrow{\;\pi\;}\; \mathcal{M}_{\phys}\,.
\end{equation}

Combining with the Cooper--Frye factorization~\eqref{eq:Factorization}, we arrive at the following structure:
\begin{equation}\label{eq:Diagram}
    \begin{array}{ccccc}
    & & G_\PGT/H_\univ[\beta,\Omega,\xi] \\[0.5em]
    & & \big\downarrow \\[0.5em]
    R_\FO & \xrightarrow{\;q\;} & \mathcal{Q}_\FO &\;\xrightarrow{\;\widetilde{\mathfrak{F}}_{\mathcal O}\;}\; & \mathcal{M}_\mathcal{O} \\[0.5em]
    & & \big\downarrow\pi \\[0.5em]
    & & \mathcal{M}_{\phys}
\end{array}
\end{equation}
The induced Cooper--Frye map $\widetilde{\mathfrak{F}}_{\mathcal O}$ factors through the horizontal arrow $q$, but it need not descend through the vertical projection $\pi$. The quotient map $q$ is universal, while the induced readout $\widetilde{\mathfrak{F}}_{\mathcal O}$ depends on the chosen observable $\mathcal O$; the vertical projection $\pi:\mathcal Q_\FO\to\mathcal M_\phys$ does not. The failure of $\widetilde{\mathfrak{F}}_{\mathcal O}$ to factor through $\pi$ is precisely the content of Buzzegoli's pseudo-gauge dependence result~\cite{Buzzegoli:2021wlg}: if such a $\overline{\mathfrak{F}}_{\mathcal O}:\mathcal{M}_\phys\to\mathcal{M}_\mathcal{O}$ existed for the polarization observable, all observables of that type would be functions of $(\beta,\Omega,\xi)$ alone, and the canonical and Belinfante pseudo-gauges would give identical predictions for $\Lambda$ polarization while they do not.

Let us spell out the relation between the original response space $R_\FO$ and the fiber $G_\PGT/H_\univ[\beta,\Omega,\xi]$. The space $R_\FO$ contains the full freeze-out surface data before quotienting: the thermodynamic Lagrange multipliers together with a choice of pseudo-gauge representative for the response operators entering $\hat K_\LE$. The map $q:R_\FO\twoheadrightarrow\mathcal Q_\FO$ forgets only those pseudo-gauge changes that lie in $H_\univ$, whereas the map $p:=\pi\circ q:R_\FO\to\mathcal M_\phys$ forgets all pseudo-gauge information and retains only the thermodynamic point $b=(\beta,\Omega,\xi)$. Thus, at fixed $b$, the relevant part of $R_\FO$ is the slice
\begin{equation}
    R_{\FO,b}:=p^{-1}(b)\subset R_\FO .
\end{equation}
Choosing one representative $r_b\in R_{\FO,b}$ gives an orbit map
\begin{equation}
    \alpha_{r_b}:\;G_\PGT\longrightarrow R_{\FO,b},\qquad
    \Phi\longmapsto \Phi\cdot r_b ,
\end{equation}
where $\Phi\cdot r_b$ denotes the response data obtained from $r_b$ by the pseudo-gauge transformation $\Phi$. The universal stabilizer at $b$ is precisely the subgroup whose action is invisible after passing to $\mathcal Q_\FO$:
\begin{equation}
    q(\Phi\cdot r_b)=q(r_b)
    \quad\Longleftrightarrow\quad
    \Phi\in H_\univ[b] .
\end{equation}
Consequently the orbit map descends to a map of coset spaces
\begin{equation}
    \bar\alpha_{r_b}:\;
    G_\PGT/H_\univ[b]\longrightarrow \pi^{-1}(b),
    \qquad
    [\Phi]\longmapsto q(\Phi\cdot r_b).
\end{equation}
When the fixed-$b$ pseudo-gauge sector of $R_\FO$ is a single $G_\PGT$-orbit, this map is an isomorphism of fibers. More generally, if $R_{\FO,b}$ has several disconnected pseudo-gauge sectors or additional non-PGT response data, $\pi^{-1}(b)$ is a union of such coset fibers, one for each orbit. In the main text we restrict to the CF pseudo-gauge sector generated from a chosen representative, so the fiber over $b$ is identified with the homogeneous space $G_\PGT/H_\univ[b]$. Strictly speaking this need not be a quotient group unless $H_\univ[b]$ is normal in $G_\PGT$; what is used here is the coset space of physically inequivalent pseudo-gauge representatives at fixed thermodynamic state.

The structure of the fiber $F_{(\beta,\Omega,\xi)} = G_\PGT/H_\univ[\beta,\Omega,\xi]$ is qualitatively different in the following two limits:
\begin{itemize}
    \item {\bf Global equilibrium.} If $\beta^\mu(x) = b^\mu + \varpi^\mu{}_\nu x^\nu$ with constant $\varpi_{\mu\nu}$, and the spin potential is equal to thermal vorticity, $\Omega_{ab} = \varpi_{ab}$, then the insertion~\eqref{eq:K_PGT} reduces, modulo boundary contributions, to a linear combination of the conserved global charges $P^\mu$, $J^{\mu\nu}$, and $Q$, which are PGT-invariant under the assumed boundary conditions. Thus $\Delta_\Phi K_\LE^{\rm GE}=0$ and $H_\univ$ encompasses essentially all of $G_\PGT$, and the fiber $F$ degenerates to a point. Each Cooper--Frye observable map $\widetilde{\mathfrak F}_{\mathcal O}$ descends all the way to $\mathcal{M}_\phys$ in this limit.
    \item {\bf Generic local equilibrium.} For $\Omega_{ab}\neq\varpi_{ab}$, the insertion~\eqref{eq:K_PGT} is a non-trivial operator and the BKM correlator $\langle \hat W;\,\Delta_\Phi\hat K_\LE\rangle_\LE$ is generically non-zero. Then $H_\univ$ is strictly smaller than $G_\PGT$, the fiber $F$ has positive dimension, and a generic observable map $\widetilde{\mathfrak F}_{\mathcal O}$ does not descend to $\mathcal{M}_\phys$.
\end{itemize}
The fiber dimension is therefore a direct measure of the local non-equilibrium spin content of the surface state.

Because $H_\univ[b]$ varies with $b$, the projection $\pi$ is not a fiber bundle in the strict (locally trivial) sense: $\mathcal{M}_\phys$ stratifies by the conjugacy type of $H_\univ[b]$, with the global-equilibrium locus $\{b : H_\univ[b] = G_\PGT\}$ as the most-degenerate stratum where $F_b$ collapses to a point. The framework as developed here uses only the existence of the projection and the variation of $F_b$, neither of which requires local triviality.

It is sometimes useful to record~\eqref{eq:Diagram} as a span of correspondences:
\begin{equation}\label{eq:Span}
    \mathcal{M}_\phys \;\xleftarrow{\;\pi\;}\; \mathcal{Q}_\FO \;\xrightarrow{\;\widetilde{\mathfrak{F}}_{\mathcal O}\;}\; \mathcal{M}_\mathcal{O}\,.
\end{equation}
The leftward and rightward arrows go between objects of genuinely different types: $\mathcal{M}_\phys$ is the space of {\em thermodynamic freeze-out states}, parameterized by Lagrange multipliers; $\mathcal{M}_\mathcal{O}$ is the space of {\em measured particle observables}, accessible at detectors; and $\mathcal{Q}_\FO$ is an intermediate object that carries both the thermodynamic state and the residual pseudo-gauge data that survive the $H_\univ$ quotient. The non-descent statement is then
\begin{equation}
    \widetilde{\mathfrak{F}}_{\mathcal O} \;\neq\; \overline{\mathfrak{F}}_{\mathcal O}\circ\pi
\end{equation}
for any candidate $\overline{\mathfrak{F}}_{\mathcal O}:\mathcal{M}_\phys\to\mathcal{M}_\mathcal{O}$.

We note that the Wigner space $\mathcal W$ introduced in~\eqref{eq:Functional} does not define a third projection out of $\mathcal Q_\FO$. It sits inside the rightward arrow of the span: for each observable, $\widetilde{\mathfrak F}_{\mathcal O}$ is obtained by composing the local-equilibrium assignment, the Wigner-function construction, and the observable functional as in~\eqref{eq:CF_composite}, and then passing to the $H_\univ$ quotient. Thus $\mathcal W$ is the intermediate phase-space representation used to evaluate observables, whereas the bases relevant for the fibration discussion below are $\mathcal M_\phys$ and $\mathcal M_\mathcal O$.

The same total space $\mathcal Q_\FO$ can therefore be viewed through two different projections. The map $\pi$ has base $\mathcal M_\phys$ and its fibers are the residual pseudo-gauge classes discussed above. The observable map $\widetilde{\mathfrak{F}}_{\mathcal O}$ has instead base $\mathcal M_\mathcal O$ (or, more precisely, its image inside $\mathcal M_\mathcal O$). If $\widetilde{\mathfrak{F}}_{\mathcal O}$ is sufficiently regular, for example a submersion onto its image in the regime considered, it may also be treated as a fibration, but its fibers are level sets of the observable map:
\begin{equation}\label{eq:ObservableLevelSet}
    \widetilde{\mathfrak{F}}_{\mathcal O}^{-1}(O)
    =
    \{\,q\in\mathcal Q_\FO\;|\;\widetilde{\mathfrak{F}}_{\mathcal O}(q)=O\,\},
    \qquad O\in {\rm Im}(\widetilde{\mathfrak{F}}_{\mathcal O}) .
\end{equation}
These measurement fibers collect all physically inequivalent points $q=[R]\in\mathcal Q_\FO$, namely $H_\univ$-classes of freeze-out data on $\Sigma_\FO$, that give the same measured observable value. In the restricted local-equilibrium sector where the assignment $\mathcal L_\LE$ in~\eqref{eq:LE_assignment} is faithful modulo $H_\univ$, one may equivalently regard these points as classes of the induced LE density operators, but the primary object remains the quotient of freeze-out data. They are generally not equal to the pseudo-gauge coset in~\eqref{eq:Fiber}: they may cut across several thermal points $(\beta,\Omega,\xi)$ and may also include residual pseudo-gauge directions to which the chosen observable is insensitive. Namely a fixed observable value can be produced by different thermal points $b=(\beta,\Omega,\xi)$, and even at fixed $b$ a particular observable may be insensitive to some residual pseudo-gauge directions beyond $H_\univ[b]$. Thus different choices of base produce different fibers. The distinguished statement of the present section is that the thermal projection $\pi$ has the pseudo-gauge orbit fiber~\eqref{eq:Fiber} under the fixed-$b$ orbit assumption, whereas the observable projection has level-set fibers determined by experimental resolving power.

This dichotomy of projections induces a corresponding two-class organization of physical observables that will be used throughout the rest of the paper. We adopt the convention that a generic ``observable'' refers to its associated map, namely the function $\widetilde{\mathfrak{F}}_\mathcal{O}:\mathcal{Q}_\FO\to\mathcal{M}_\mathcal{O}$ that assigns to each equivalence class $[R]\in\mathcal{Q}_\FO$ the value of the observable on the corresponding physical state. The image $\widetilde{\mathfrak{F}}_\mathcal{O}([R])\in\mathcal{M}_\mathcal{O}$ is the {\em value} (a number or small tuple); the map $\widetilde{\mathfrak{F}}_\mathcal{O}$ itself is the {\em observable} in this geometric sense. Since every physical observable is an expectation value $\langle\hat A\rangle_\LE = \mathrm{Tr}(\hat\rho_\LE\hat A)$ and $\hat\rho_\LE$ is invariant on $H_\univ$-orbits, every such map is automatically well-defined on $\mathcal{Q}_\FO$. We call $\widetilde{\mathfrak{F}}_\mathcal{O}:\mathcal{Q}_\FO\to\mathcal{M}_\mathcal{O}$ a {\bf base observable} if it further descends through $\pi$ to a single-valued function on $\mathcal{M}_\phys$, i.e.\ takes the same value at all points of a $\pi$-fiber, and a {\bf fiber observable} otherwise. The sharp criterion is
\begin{equation}\label{eq:base_vs_fiber}
    \widetilde{\mathfrak{F}}_\mathcal{O}\;\text{is a base observable} \;\Longleftrightarrow\; \widetilde{\mathfrak{F}}_\mathcal{O}\;\text{is constant on each fiber}\;\pi^{-1}(b),\; b\in\mathcal{M}_\phys\,.
\end{equation}
In the heavy-ion polarization context, the total baryon charge $\hat Q$, the total energy-momentum $\hat P^\nu$, and kinematic vorticity moments computed from the reconstructed flow profile are base observables; the local $\Lambda$ polarization $S^\mu(p)$~\cite{Buzzegoli:2021wlg, Liu:2021nyg, Becattini:2013fla}, the vector-meson spin alignment $\rho_{00}$~\cite{STAR:2022fan, Sheng:2025cjk}, and the axial vortical conductivity~\cite{Buzzegoli:2021wlg} are fiber observables. Note that the total angular momentum flux $\hat J^{\mu\nu}_\Sigma = \int_\Sigma d\Sigma_\lambda\,\hat J^{\lambda,\mu\nu}$ is PGT-invariant on $\Sigma_\FO\cong\mathbb{R}^3$ and hence a base observable, whereas the global $\Lambda$ polarization $\langle P_y\rangle$, despite also being an integrated quantity, is not: $\langle P_y\rangle$ is constructed from the Wigner function via the momentum-dependent spin-vector projection~\eqref{eq:S_LE}, weighted by $n_F(1-n_F)$ Fermi--Dirac factors and the Cooper--Frye flux $d\Sigma\!\cdot\!p$, and therefore probes a momentum-resolved component of the spin density rather than the total angular momentum charge. It is this momentum-space kernel that prevents the PGT-dependent local spin densities from canceling in the integral, leaving $\langle P_y\rangle$ sensitive to the pseudo-gauge fiber. The remainder of the paper develops the consequences of this dichotomy.

\section{Predictions from the Fibration}\label{sec:predictions}

Having established the fibration structure of equivalence relations in section~\ref{sec:fibration} and the resulting base/fiber dichotomy of observables, we now extract concrete, falsifiable consequences for the physics of polarization observables at freeze-out. The strategy is uniform throughout: each prediction is read off from the geometry of the diagram~\eqref{eq:Diagram} together with the master formula~\eqref{eq:Delta_Ophys}, with no further dynamical input beyond the local-equilibrium assumption already made in section~\ref{sec:setup}. The phenomenological output is therefore framework-level rather than numerical: it organizes how observables are constrained within the local-equilibrium and fixed-multiplier setup, identifies independent PG-sensitive moments and cross-observable consistency relations, and isolates the inputs (the spin potential profile $\Omega_{ab}(x)$ on $\Sigma_\FO$ and any additional local response data) needed to convert these qualitative predictions into numerical ones.

In section~\ref{sec:counting} we define the fiber dimension carefully in both its local pointwise form $d_F(x)$ and its experimentally relevant family-restricted form $d_F^\mathcal{F}$, prove that both vanish at global equilibrium, and use them to upper-bound the number of independent fiber observables at a given thermodynamic state. In section~\ref{sec:consistency} we show that two fiber observables measured in the same kinematic bin must lie on a specific subvariety of the product observable space, computable from a single RSH input, and then apply this statement to the tension between $\Lambda$ hyperon polarization and $\phi$-meson spin alignment and to the extended field-correlation response sector proposed in~\cite{Sheng:2022wsy}. Each prediction is derived directly from the geometry of the diagram~\eqref{eq:Diagram}; none requires the explicit form of the Wigner function or specific assumptions beyond local equilibrium.

\subsection{Counting Bound on Independent PG-Sensitive Observables}\label{sec:counting}

Naively one would like to define the fiber dimension as $\dim_\mathbb{R}\bigl(G_\PGT/H_\univ[\beta,\Omega,\xi]\bigr)$ and count it as the maximal number of independent PG-sensitive observables. This is not directly meaningful, because both $G_\PGT$ and $H_\univ$ are infinite-dimensional function spaces: the pseudo-gauge generator $\Phi^{\lambda,\mu\nu}(x)$ is an antisymmetric-tensor-valued field on spacetime, with a continuum of independent components at every point of $\Sigma_\FO$. The literal coset dimension is therefore of the form $\infty - \infty$ and requires a regularization to extract a finite, physically meaningful number. There are two natural ways to do this, both relevant for distinct phenomenological questions.

\paragraph{(i) Local (pointwise) fiber dimension.} At a single point $x\in\Sigma_\FO$, the values $\Phi^{\lambda,\mu\nu}(x)$ form a finite-dimensional vector space. The free index $\lambda$ ranges over $D=4$ values while the antisymmetric pair $\mu\nu$ ranges over $D(D-1)/2 = 6$ values, so the local PGT generator at $x$ lives in
\begin{equation}\label{eq:Phi_dim}
    V_\PGT^{(x)} \;:=\; \bigl\{\Phi^{\lambda,\mu\nu}(x) \,:\, \Phi^{\lambda,\mu\nu}(x) = -\Phi^{\lambda,\nu\mu}(x)\bigr\}\,,\qquad \dim_\mathbb{R} V_\PGT^{(x)} \;=\; 24\,.
\end{equation}
To regularize the codimension of $H_\univ$ inside $G_\PGT$ at the pointwise level, we use the evaluation map
\begin{equation}\label{eq:Evaluation}
    \mathrm{ev}_x:\;G_\PGT\longrightarrow V_\PGT^{(x)}\,,\qquad \Phi\longmapsto \Phi(x)\,,
\end{equation}
and its restriction to the stabilizer $\mathrm{ev}_x\big|_{H_\univ}: H_\univ\to V_\PGT^{(x)}$. The image of this restriction
\begin{equation}\label{eq:V_Huniv}
    \begin{split}
    	V_{H_\univ}^{(x)}[\beta,\Omega,\xi] \;&:=\; \mathrm{ev}_x\bigl(H_\univ[\beta,\Omega,\xi]\bigr) \;\\
    	&=\; \bigl\{\Phi(x)\in V_\PGT^{(x)} \,:\, \exists\,\tilde\Phi\in H_\univ\,,\;\tilde\Phi(x) = \Phi(x)\bigr\} \;\subset\; V_\PGT^{(x)}
    \end{split}
\end{equation}
collects all pointwise values that can be realized by some global pseudo-gauge transformation in the universal stabilizer. We define the {\bf local fiber dimension} at $x$ as
\begin{equation}\label{eq:dFlocal}
    d_F(x;\beta,\Omega,\xi) \;:=\; \dim_\mathbb{R}\bigl(V_\PGT^{(x)} \big/ V_{H_\univ}^{(x)}[\beta,\Omega,\xi]\bigr)\,, \qquad 0 \;\leq\; d_F(x;\beta,\Omega,\xi) \;\leq\; 24\,.
\end{equation}
Operationally, $d_F(x)$ counts those local PG directions at $x$ that cannot be neutralized by any global completion lying in $H_\univ$: it is the dimension of the obstruction, at $x$, to extending a given pointwise variation to a globally observable-invariant pseudo-gauge transformation. However, there is an important caveat: because $H_\univ$ is defined by the surface-integral condition~\eqref{eq:Def_Huniv} rather than a pointwise constraint, a global stabilizer element can in principle realize arbitrary local values at $x$ while compensating elsewhere on $\Sigma_\FO$. The image $V_{H_\univ}^{(x)}$ may therefore be larger than one would guess from the local form of the stabilizer condition, and $d_F(x)$ should accordingly be regarded as a heuristic indicator of local PG sensitivity rather than a sharp count. The family-restricted fiber dimension $d_F^\mathcal{F}$ defined below does not suffer from this subtlety, since it operates within a finite-dimensional family where the global and local conditions are directly comparable, and is the operationally preferred quantity.

\paragraph{Global-equilibrium check: $d_F(x) = 0$.} The Proposition of section~\ref{sec:GE_full} establishes that at global equilibrium, with $\beta^\mu$ Killing, $\Omega_{ab}=\varpi_{ab}$, and constant chemical potential, the universal stabilizer saturates the entire pseudo-gauge group:
\begin{equation}\label{eq:Huniv_GE_reminder}
    H_\univ[\beta_{\rm Killing},\,\Omega = \varpi] \;=\; G_\PGT\,.
\end{equation}
The evaluation map then has $H_\univ = G_\PGT$ as its full domain, and the image~\eqref{eq:V_Huniv} is the full image of $\mathrm{ev}_x$ on $G_\PGT$. The latter is all of $V_\PGT^{(x)}$, because for any prescribed value $\Phi_0\in V_\PGT^{(x)}$ one may choose a smooth bump function $\chi(y)$ with $\chi(x) = 1$ and support contained in any prescribed neighborhood of $x$, and set $\tilde\Phi^{\lambda,\mu\nu}(y) = \chi(y)\,\Phi_0^{\lambda,\mu\nu}$; this $\tilde\Phi$ lies in $G_\PGT$ (and in $H_\univ = G_\PGT$ at this state) and satisfies $\tilde\Phi(x) = \Phi_0$. Thus
\begin{equation}\label{eq:Vuniv_GE}
    V_{H_\univ}^{(x)}[\beta_{\rm Killing},\,\Omega = \varpi] \;=\; V_\PGT^{(x)}\quad\text{for every}\; x\in\Sigma_\FO\,,
\end{equation}
and from~\eqref{eq:dFlocal}
\begin{equation}\label{eq:dF_GE}
    d_F(x;\beta_{\rm Killing},\,\Omega = \varpi,\,\xi=\mathrm{const}) \;=\; 0 \quad \text{for every}\; x\in\Sigma_\FO\,.
\end{equation}
The local fiber dimension therefore vanishes pointwise at global equilibrium, consistently with the global statement that the fiber $G_\PGT/H_\univ$ of the fibration~\eqref{eq:Fibration} degenerates to a point. There are no local PG-sensitive directions in this limit. This is the precise sense in which the geometry of the fibration realizes the Belinfante--Rosenfeld content at the level of local data.

\paragraph{Local-equilibrium case.} Away from global equilibrium, $H_\univ\subsetneq G_\PGT$, and the image~\eqref{eq:V_Huniv} is generically a strict subspace of $V_\PGT^{(x)}$, so $d_F(x)>0$ is expected at generic points, subject to the caveat above. The Belinfante--canonical comparison made explicit in eq.~\eqref{eq:dF_BC_value} below provides the simplest one-dimensional realization at the family-restricted level.

\paragraph{(ii) Family-restricted fiber dimension.} In experimental practice one does not vary $\Phi^{\lambda,\mu\nu}(x)$ over all $24\,\mathrm{vol}(\Sigma_\FO)$ functional directions, but compares predictions from a finite list of physically motivated pseudo-gauge choices: Belinfante, canonical, GLW, HW, and so on~\footnote{These are four standard pseudo-gauges in the spin-tensor literature~\cite{Buzzegoli:2021wlg, Speranza:2020ilk}. The {\em canonical} pseudo-gauge $(\hat T_\can, \hat S_\can)$ is the pair derived from Noether's theorem applied to the Dirac Lagrangian, with $\hat S_\can^{\lambda,\mu\nu}$ totally antisymmetric. The {\em Belinfante} pseudo-gauge $(\hat T_\Bel, 0)$ is obtained from the canonical pair by the symmetrization procedure of~\cite{Belinfante:1940, Rosenfeld:1940}, which yields a symmetric stress tensor at the cost of setting the spin current to zero. The {\em de Groot--van Leeuwen--van Weert} (GLW) pseudo-gauge~\cite{deGroot:1980} is the one in which the spin current is constructed so that the corresponding Wigner-function representation matches the classical kinetic-theory expansion of relativistic transport. The {\em Hilgevoord--Wouthuysen} (HW) pseudo-gauge~\cite{Hilgevoord:1963} provides a conserved spin operator $\hat{\vec S}_{\rm HW}$ for the free Dirac electron, i.e.\ one that commutes with the free Dirac Hamiltonian. It is the relativistic generalization of the non-relativistic Foldy--Wouthuysen mean-spin operator obtained by the Foldy--Wouthuysen transformation~\cite{Foldy:1950}, which block-diagonalizes the Dirac Hamiltonian and isolates the positive-energy subspace; the orbital angular momentum complementary to $\hat{\vec S}_{\rm HW}$ is correspondingly based on the FW mean-position operator $\hat{\vec X}_{\rm FW}$ rather than on the standard Dirac position $\hat{\vec x}$. Each of these four corresponds to a specific element of $G_\PGT$ and produces, in general, distinct polarization predictions; explicit canonical-vs-Belinfante-vs-GLW-vs-HW differences are computed in~\cite{Buzzegoli:2021wlg}.}. Each such choice corresponds to a specific element $\Phi_{(a)} \in G_\PGT$, and the relevant family is a finite-dimensional linear span
\begin{equation}\label{eq:F_def}
    \mathcal{F} \;:=\; \mathrm{span}_\mathbb{R}\bigl\{\Phi_{(1)},\,\Phi_{(2)},\,\ldots,\,\Phi_{(N)}\bigr\} \;\subset\; G_\PGT\,, \qquad \dim_\mathbb{R}\mathcal{F} \;=\; N\,.
\end{equation}
The {\bf family-restricted fiber dimension} is then the codimension of $H_\univ$ inside this finite family:
\begin{equation}\label{eq:dFfamily}
    d_F^\mathcal{F}[\beta,\Omega,\xi] \;:=\; \dim_\mathbb{R}\bigl(\mathcal{F} \big/ (\mathcal{F}\cap H_\univ[\beta,\Omega,\xi])\bigr)\,, \qquad 0\;\leq\; d_F^\mathcal{F} \;\leq\; N\,.
\end{equation}
The two integers in~\eqref{eq:dFlocal} and~\eqref{eq:dFfamily} answer different physical questions: $d_F(x)$ measures the local richness of PG ambiguity per spacetime point on $\Sigma_\FO$, whereas $d_F^\mathcal{F}$ measures how many of the experimentally accessible PG-comparison directions actually distinguish observables at the given thermodynamic state.

With both notions in hand, the counting prediction can now be stated precisely as follows:

\noindent{\bf Local counting bound.} At each $x\in\Sigma_\FO$, the local fiber dimension $d_F(x)$ provides an upper bound on the number of independent local PG-sensitive moments that fiber observables can build at that point. Because $H_\univ$ is a global rather than pointwise condition, this bound should be understood as heuristic rather than sharp.

\noindent{\bf Family-restricted counting bound.} For a given finite family $\mathcal{F}$ of pseudo-gauge choices, there exist at most $d_F^\mathcal{F}$ functionally independent fiber observables whose PG-class dependence is detectable by varying $\Phi$ within $\mathcal{F}$. If experiments compare $N$ pseudo-gauge choices but report fewer than $d_F^\mathcal{F}$ independent PG-sensitive observables across the comparison, the framework predicts that additional independent ones can exist, and identifies the candidate unseen ones with the remaining $d_F^\mathcal{F}$ directions of $\mathcal{F}\big/(\mathcal{F}\cap H_\univ)$.

\paragraph{Application to the Buzzegoli analysis.} The Belinfante--canonical comparison of~\cite{Buzzegoli:2021wlg} corresponds to the simplest non-trivial family
\begin{equation}\label{eq:F_BC}
    \mathcal{F}_{\Bel\text{--}\can} \;=\; \mathrm{span}_\mathbb{R}\bigl\{\hat S_\can^{\lambda,\mu\nu}\bigr\}\,,\qquad N = 1\,,
\end{equation}
i.e.\ the one-parameter line of pseudo-gauges $\Phi_\alpha = \alpha\hat S_\can$ interpolating from the canonical representative at $\alpha = 0$ to the Belinfante representative at $\alpha = 1$. At first order in thermodynamic gradients, the master formula~\eqref{eq:K_PGT} on this family reduces, as derived in section~\ref{sec:BC_test}, to
\begin{equation}
    \Delta_{\Phi_\alpha}\hat K_\LE \;\propto\; \alpha \int_{\Sigma_\FO} d\Sigma_\mu\,(\Omega_{ab}-\varpi_{ab})\,\hat S_\can^{\mu,ab}\,,
\end{equation}
which, for generic spin-density configurations, equals $c_\Phi\mathbf{1}$ if and only if $\Omega_{ab} = \varpi_{ab}$ on $\Sigma_\FO$.\footnote{The genericity assumption is that no non-trivial surface-integrated linear combination of $\hat S_\can^{\mu,ab}$ weighted by $(\Omega_{ab}-\varpi_{ab})$ reduces to a multiple of the identity when $\Omega_{ab}\neq\varpi_{ab}$.} Hence
\begin{equation}\label{eq:dF_BC_value}
    d_F^{\mathcal{F}_{\Bel\text{--}\can}}[\beta,\Omega,\xi] \;=\; \begin{cases} 0 & \text{if } \Omega_{ab} = \varpi_{ab}\text{ on }\Sigma_\FO\text{ (in particular at global equilibrium)},\\ 1 & \text{otherwise}. \end{cases}
\end{equation}
The single direction in $\mathcal{F}_{\Bel\text{--}\can}/(\mathcal{F}_{\Bel\text{--}\can}\cap H_\univ)$ corresponds to the $(\varpi-\Omega)$-weighted moment that Buzzegoli computes explicitly in his Eqs.~(7)--(9).

Enlarging $\mathcal{F}$ to include additional pseudo-gauges, such as the GLW, HW, or Hilgevoord--Wouthuysen choices of~\cite{Buzzegoli:2021wlg}, or the most general PGT with a non-trivial spin superpotential $Z^{ab,\mu\rho}$ in the sense of~\cite{Speranza:2020ilk}, yields a strictly larger $\mathcal{F}$, larger $N$, and generically a strictly larger $d_F^\mathcal{F}$. Each new independent direction in $\mathcal{F}/(\mathcal{F}\cap H_\univ)$ can support a new independent fiber observable beyond the canonical--Belinfante difference, provided the available observable family separates that vertical direction. This is the precise content of the counting bound as a phenomenological discovery target.

\subsection{Cross-Observable Consistency Relations}\label{sec:consistency}

A central structural consequence of the framework is that fiber-observable values measured on the same physical state must arise from a single point of $\mathcal{Q}_\FO$, with ``physical state'' understood throughout in the sense fixed in section~\ref{sec:fibration}, i.e.\ as a point $[R]\in\mathcal{Q}_\FO$ specifying both the thermal Lagrange multipliers and the residual pseudo-gauge representative. The reason becomes transparent once one retraces the chain of section~\ref{sec:setup}: for any observable, its value $\mathcal{O}\in\mathcal{M}_\mathcal{O}$ on a given state is computed as a functional of the Wigner function,
\begin{equation}\label{eq:O_value_def}
    \mathcal{O} \;=\; \mathfrak{f}_\mathcal{O}\bigl[W(x,p)\bigr]\,,
\end{equation}
with $W(x,p) = \mathrm{Tr}(\hat\rho_\LE \hat W(x,p))$ depending on the freeze-out data only through the LE density operator $\hat\rho_\LE$. As recalled in section~\ref{sec:fibration}, an equivalence class $[R]\in\mathcal{Q}_\FO$ specifies precisely the combination of thermal Lagrange multipliers and residual pseudo-gauge representative needed to fix $\hat\rho_\LE$, with two points of $R_\FO$ identified in the quotient whenever they are related by an element of $H_\univ$ via~\eqref{eq:Univ_Stab}. In the faithful local-equilibrium sector, or after quotienting by any residual non-PGT redundancies, the assignment $[R]\mapsto\hat\rho_\LE$ may be identified with the space of physically distinct LE density operators; in general, $\mathcal{Q}_\FO$ is the universal PGT quotient on which all admissible Cooper--Frye observables are well-defined. In either case,~\eqref{eq:O_value_def} factors uniquely through $\mathcal{Q}_\FO$: the observable value $\mathcal{O}\in\mathcal{M}_\mathcal{O}$ depends on the freeze-out data only through $[R]$ via a well-defined map
\begin{equation}\label{eq:Otilde_def}
    \widetilde{\mathfrak{F}}_\mathcal{O}:\;\mathcal{Q}_\FO \longrightarrow \mathcal{M}_\mathcal{O}\,, \qquad \mathcal{O}\;=\;\widetilde{\mathfrak{F}}_\mathcal{O}\bigl([R]\bigr)\,.
\end{equation}

The substantive content of the statement is then the following: distinct observable values $\mathcal{O}_1, \mathcal{O}_2, \ldots\in\mathcal{M}_\mathcal{O}^{(1)},\mathcal{M}_\mathcal{O}^{(2)},\ldots$ on the same physical state are values of different functions $\widetilde{\mathfrak{F}}_{\mathcal O_1}, \widetilde{\mathfrak{F}}_{\mathcal O_2}, \ldots$ evaluated at the same argument $[R]$. The joint observable map cannot ``decouple'' across observables, because all such functions are induced by the same local-equilibrium density operator $\hat\rho_\LE \equiv [R]$. For two such observable maps $\widetilde{\mathfrak{F}}_{\mathcal O_a} : \mathcal{Q}_\FO \to \mathcal{M}_\mathcal{O}^{(a)}$ with $a=1,2$, the simultaneous measured values $(\mathcal{O}_1,\mathcal{O}_2)$ in a single kinematic bin must therefore lie on the image of $\mathcal{Q}_\FO$ inside the product space:
\begin{equation}\label{eq:Graph}
    \bigl(\mathcal{O}_1,\,\mathcal{O}_2\bigr)\;\in\;\mathrm{Im}\bigl(\widetilde{\mathfrak{F}}_{\mathcal O_1}\times\widetilde{\mathfrak{F}}_{\mathcal O_2}\bigr)
    \;=\;\Bigl\{\bigl(\widetilde{\mathfrak{F}}_{\mathcal O_1}([R]),\,\widetilde{\mathfrak{F}}_{\mathcal O_2}([R])\bigr)\,:\,[R]\in\mathcal{Q}_\FO\Bigr\}
    \;\subset\;\mathcal{M}_\mathcal{O}^{(1)}\times\mathcal{M}_\mathcal{O}^{(2)}\,.
\end{equation}
When $\widetilde{\mathfrak{F}}_{\mathcal O_1}$ is injective onto its image,~\eqref{eq:Graph} is the graph of the composition $\widetilde{\mathfrak{F}}_{\mathcal O_2}\circ\widetilde{\mathfrak{F}}_{\mathcal O_1}^{-1}$. The image is computable from any realistic relativistic spin hydrodynamics simulation that specifies $(\beta,\Omega,\xi)|_{\Sigma_\FO}$ together with a pseudo-gauge choice.

The non-triviality of this statement is best appreciated by contrasting it with what {\em a priori} would be the case. If $\widetilde{\mathfrak{F}}_{\mathcal O_1}$ and $\widetilde{\mathfrak{F}}_{\mathcal O_2}$ were unrelated functions of the freeze-out data, the joint pair $(\mathcal{O}_1,\mathcal{O}_2)$ could in principle take values anywhere in the full product space $\mathcal{M}_\mathcal{O}^{(1)}\times\mathcal{M}_\mathcal{O}^{(2)}$, with each coordinate constrained only by its individual range. The fibration structure removes this freedom: both maps share the common domain $\mathcal{Q}_\FO$ and are evaluated at the same argument $[R]$, so the realizable joint values are confined to the image of the single joint evaluation map
\begin{equation}\label{eq:JointEval}
    \widetilde{\mathfrak{F}}_{\mathcal O_1}\times\widetilde{\mathfrak{F}}_{\mathcal O_2}:\;\mathcal{Q}_\FO \longrightarrow \mathcal{M}_\mathcal{O}^{(1)}\times\mathcal{M}_\mathcal{O}^{(2)}\,,\qquad [R]\longmapsto\bigl(\widetilde{\mathfrak{F}}_{\mathcal O_1}([R]),\,\widetilde{\mathfrak{F}}_{\mathcal O_2}([R])\bigr)\,.
\end{equation}
Equivalently, the marginal ranges of $\widetilde{\mathfrak{F}}_{\mathcal O_1}$ and $\widetilde{\mathfrak{F}}_{\mathcal O_2}$ are necessary but not sufficient constraints on the joint measurement: simultaneous measurement of two fiber observables in the same kinematic bin carries strictly more information than the union of the two marginal measurements. A quantitative version of this statement, in terms of a codimension bound on $\mathrm{Im}(\widetilde{\mathfrak{F}}_{\mathcal O_1}\times\widetilde{\mathfrak{F}}_{\mathcal O_2})$ inside the product, is delicate as written: $\mathcal{Q}_\FO$ is in general infinite-dimensional, since it includes all field profiles $(\beta,\Omega,\xi)|_{\Sigma_\FO}$ together with the residual pseudo-gauge data, so the image can a priori fill an open subset of a finite-dimensional product such as $\mathbb{R}^2$. Strict-codimension content arises after one restricts to an experimentally accessible finite-parameter slice $S\subset\mathcal{Q}_\FO$, which corresponds to a specific RSH simulation together with a chosen pseudo-gauge family $\mathcal{F}$ and a small number of scanned parameters. On this slice $\mathrm{Im}(\widetilde{\mathfrak{F}}_{\mathcal O_1}\times\widetilde{\mathfrak{F}}_{\mathcal O_2}|_S)$ has dimension at most $\dim S$ and is generically a low-dimensional curve or surface in the product. It is on this curve or surface that the joint experimental pair $(\mathcal{O}_1^{\rm exp},\mathcal{O}_2^{\rm exp})$ must lie.

\paragraph{Operational test.} The dimension of the accessible slice $S\subset\mathcal{Q}_\FO$ is itself controlled by physically meaningful choices made in setting up $\hat K_\LE$, and provides a sharp diagnostic of those choices that is internal to the fibration framework. The simplest non-trivial instance is the following. Consider two real-valued fiber observables $\mathcal{O}_1,\mathcal{O}_2$, so that the product is $\mathbb{R}^2$. Fix the thermal point $b=(\beta,\Omega,\xi)\in\mathcal{M}_\phys$ to whatever value a chosen relativistic spin-hydrodynamics (RSH) simulation~\footnote{By an ``RSH simulation'' we mean a numerical evolution of the QGP that treats the spin degrees of freedom on equal footing with the standard fluid variables, producing on the freeze-out hypersurface $\Sigma_\FO$ the full set of fields $(\beta^\mu, T, u^\mu, \xi, \Omega_{ab})$ needed to construct $\hat K_\LE$. Several explicit formulations exist, differing in their treatment of spin relaxation and dissipative transport but sharing the same overall structure~\cite{Florkowski:2017ruc, Florkowski:2018fap, Hongo:2021ona, Gallegos:2021bzp, Gallegos:2022jow}. The output of an RSH simulation is the input to the Cooper--Frye chain of section~\ref{sec:setup}.} produces in a given kinematic bin~\footnote{In heavy-ion experiments, polarization and spin-alignment observables are measured differentially in a multi-dimensional phase space whose axes are the transverse momentum $p_T$, the rapidity $y$ (or pseudorapidity $\eta$), the azimuthal angle $\phi$ relative to the reaction plane, the collision centrality class, and the center-of-mass energy per nucleon pair $\sqrt{s_{NN}}$. A \emph{kinematic bin} denotes a small region of this space (e.g., 20--50\% centrality, $1.0<p_T<2.0$~GeV/$c$, $|y|<1$, $\sqrt{s_{NN}}=200$~GeV); within such a bin the data are averaged over events to yield a single observable value. Cross-observable consistency at fixed kinematic bin compares two such bin-averaged observables that probe (approximately) the same underlying freeze-out state.}, and restrict the pseudo-gauge family $\mathcal{F}$ to the one-parameter canonical-to-Belinfante interpolation $\mathcal{F}_{\Bel\text{--}\can}=\mathrm{span}_\mathbb{R}\{\hat S_\can\}$ introduced in section~\ref{sec:GE_full} and analyzed in~\eqref{eq:F_BC}--\eqref{eq:dF_BC_value} above. As recorded in~\eqref{eq:dF_BC_value}, at a generic local-equilibrium point with $\Omega\neq\varpi$ this family contributes a single direction to the fiber, $d_F^{\mathcal{F}_{\Bel\text{--}\can}}=1$, so the accessible slice $S\subset\pi^{-1}(b)\subset\mathcal{Q}_\FO$ is itself one-dimensional, traced out by the single superpotential direction $\Phi_{\rm B}$ in the fiber $G_\PGT/H_\univ[b]$, and its image under $\widetilde{\mathfrak{F}}_{\mathcal O_1}\times\widetilde{\mathfrak{F}}_{\mathcal O_2}$ is therefore a curve in $\mathbb{R}^2$. The cross-observable consistency relation~\eqref{eq:Graph} is the geometric statement that the simultaneous experimental pair $(\mathcal{O}_1^{\rm exp},\mathcal{O}_2^{\rm exp})$ in the same kinematic bin must lie on this curve. An on-curve measurement is consistent with both the chosen RSH input and the minimal response sector underlying $\hat K_\LE$; an off-curve measurement signals one of three possibilities: (i)~the minimal response sector underlying $\hat K_\LE$ is too narrow and must be enlarged to an extended generator $\hat K_\LE^{\rm ext}$ incorporating further fiber-active responses beyond the minimal $(\hat T,\hat S,\hat j)$, lifting the dimension of $S$ and of $\mathrm{Im}(\widetilde{\mathfrak{F}}_{\mathcal O_1}\times\widetilde{\mathfrak{F}}_{\mathcal O_2}|_S)$ so that the image can broaden from a curve to a two-dimensional region of $\mathbb{R}^2$ in which the experimental point can sit; (ii)~the RSH dynamics generating $(\beta,\Omega,\xi)$ on $\Sigma_\FO$ is inaccurate, i.e.\ the simulation input chosen to fix $b$ is itself wrong; or (iii)~the two observables in fact probe distinct underlying states because of decay or detector effects not captured by the fibration. The last two are external to the framework and concern the inputs and the experimental reconstruction respectively, while (i) is internal to it and produces the structural prediction that enlarging $\hat K_\LE$ is required; the concrete construction of $\hat K_\LE^{\rm ext}$ will be carried out in the heavy-ion specialization below. Standard heavy-ion analyses do not impose this constraint because they do not articulate the fibration structure. The framework therefore provides a sharper interpretation tool than merely fitting each observable separately and comparing.

\paragraph{Two fiber observables in the heavy-ion context.} The most experimentally accessible pair of fiber observables on $\mathcal{Q}_\FO$ in heavy-ion collisions are the global $\Lambda$ hyperon polarization $\langle P_y\rangle$ and the vector-meson spin alignment $\rho_{00}^\phi$. The first is defined as the projection of the mean spin polarization four-vector $S^\mu(p)$ of the $\Lambda$ onto the direction $\hat y$ of the global orbital angular momentum (perpendicular to the reaction plane), integrated over the produced-$\Lambda$ momentum distribution. Experimentally it is reconstructed from the angular distribution of the daughter proton in the weak decay $\Lambda\to p\pi^-$,
\begin{equation}\label{eq:Lambda_angdist}
    \frac{dN}{d\cos\theta^*} \;\propto\; 1 + \alpha_H\,\langle P_y\rangle\,\cos\theta^*\,,
\end{equation}
where $\theta^*$ is the polar angle of the proton with respect to $\hat y$ in the $\Lambda$ rest frame and $\alpha_H\simeq 0.732$ is the $\Lambda$ weak decay parameter. The mean spin vector $S^\mu(p)$ is obtained from the Wigner function $W(x,p)$ at freeze-out~\cite{Buzzegoli:2021wlg, Becattini:2013fla, Liu:2021nyg}; because $S^\mu(p)$ depends on the spin potential $\Omega_{ab}$ in addition to the thermal vorticity, $\langle P_y\rangle$ is a fiber observable in our sense, not a base observable.

The second is the $(0,0)$ diagonal entry of the $3\times3$ spin density matrix $\rho^{(s)}_{ij}$ of the spin-$1$ $\phi$ meson, with the spin-quantization axis chosen along the global orbital angular momentum direction $\hat y$. Experimentally $\rho_{00}^\phi$ is reconstructed from the angular distribution of the $\phi\to K^+K^-$ decay products,
\begin{equation}\label{eq:phi_angdist}
    \frac{dN}{d\cos\theta^*} \;\propto\; (1-\rho_{00}^\phi) + (3\rho_{00}^\phi-1)\cos^2\theta^*\,,
\end{equation}
where $\theta^*$ is the polar angle of a daughter kaon with respect to $\hat y$ in the $\phi$ rest frame. For an unpolarized emission, equal populations of the three spin states give $\rho_{00}^\phi = 1/3$; any deviation measures tensor spin alignment of the $\phi$~\cite{Yang:2017sdk, STAR:2022fan, Sheng:2025cjk}. In the minimal local-equilibrium closure used above, the spin density matrix is treated as a functional of the freeze-out phase-space and spin-density data induced by the same $\hat\rho_\LE$, evaluated in the spin-$1$ sector of $\phi$ formation, so $\rho_{00}^\phi$ is likewise a fiber observable on $\mathcal{Q}_\FO$.

Within this minimal closure, both $\langle P_y\rangle$ and $\rho_{00}^\phi$ are read out from the common freeze-out state $\hat\rho_\LE$, using the phase-space or spin-density representation appropriate to each hadron, and therefore descend simultaneously to functions on $\mathcal{Q}_\FO$. A single freeze-out state then corresponds to a single point $[R]\in\mathcal{Q}_\FO$ and a single pair $(\langle P_y\rangle,\rho_{00}^\phi)\in\mathcal{M}_\mathcal{O}^{(\Lambda)}\times\mathcal{M}_\mathcal{O}^{(\phi)}$, which by~\eqref{eq:Graph} must lie on the image of $\mathcal{Q}_\FO$ under the joint map $\widetilde{\mathfrak{F}}_\Lambda\times\widetilde{\mathfrak{F}}_\phi$. The cross-observable consistency relation is therefore the geometric statement that the two-dimensional joint measurement is constrained to lie on this image rather than anywhere in the product space.

\paragraph{Phenomenological context: $\Lambda$ hyperon polarization versus $\phi$-meson spin alignment.} The current experimental status of the two observables is suggestive, although the quoted published numbers below are not yet strictly-speaking a same-bin joint measurement. STAR measured $\langle P_y\rangle = 0.277 \pm 0.040\,(\text{stat}) \pm 0.049\,(\text{sys})\,\%$ for the $\Lambda$ in Au+Au collisions at $\sqrt{s_{NN}} = 200$~GeV~\cite{STAR:2018gyt}, and somewhat larger values, of order $1\%$ to a few percent, at the lower beam-energy-scan energies~\cite{STAR:2017ckg}; these magnitudes are broadly consistent with vorticity-dominated relativistic spin-hydrodynamic predictions. The same collaboration subsequently measured the $\phi$-meson alignment averaged over the BES energy window $\sqrt{s_{NN}} = 11.5$--$62$~GeV, obtaining~\cite{STAR:2022fan}
\begin{equation}\label{eq:rho00_value}
    \rho_{00}^\phi - \tfrac13 \;=\; 0.018 \pm 0.002\,(\text{stat}) \pm 0.002\,(\text{syst})\,,
\end{equation}
while the corresponding $K^{*0}$ alignment is consistent with $1/3$ in the same kinematic window. The conventional vorticity-coalescence prediction is schematic, obtained by combining two standard ingredients. The first is the coalescence formula for the vector-meson spin density matrix in terms of the constituent quark and antiquark polarizations $P_q,P_{\bar q}$~\cite{Liang:2004xn, Yang:2017sdk},
\begin{equation}\label{eq:rho00_coalescence}
    \rho_{00} \;=\; \frac{1-P_q P_{\bar q}}{3+P_q P_{\bar q}}
    \;=\; \frac{1}{3} \,-\, \frac{4}{9}\,P_q P_{\bar q} \,+\, O\!\bigl((P_q P_{\bar q})^2\bigr)\,,
\end{equation}
the second equality being the small-polarization expansion appropriate to the vorticity-driven regime~\footnote{The expansion is justified numerically: at typical QGP temperatures $T\sim 150\text{--}300\,\mathrm{MeV}$ and angular-velocity scales extracted from the most-vortical-fluid measurements~\cite{STAR:2017ckg}, the Becattini-type estimate gives a constituent-quark polarization $P_q\sim \omega/(2T) \sim 5\times 10^{-3}\text{--}5\times 10^{-2}$, hence $|P_q P_{\bar q}|\sim P_q^2\sim 2.5\times 10^{-5}\text{--}2.5\times 10^{-3}$. The series for $(1-x)/(3+x)$ has radius of convergence $|x|<3$, so the $O((P_q P_{\bar q})^2)$ remainder is utterly negligible at any phenomenologically relevant accuracy within the vorticity-only baseline that the parabola~\eqref{eq:naive_curve} is designed to represent.} with the leading deviation from $1/3$ negative because parallel quark-antiquark polarizations preferentially populate the $m=\pm1$ vector states. The second ingredient is that, at leading order in thermal vorticity, both the $\Lambda$ polarization and the constituent-quark polarization are linear in $\varpi$. On the $\Lambda$ side, in the freeze-out-integrated form used in heavy-ion phenomenology, we have the on-shell mean spin-polarization four-vector of a Dirac particle on a freeze-out hypersurface as~\cite{Becattini:2013fla, Liu:2021nyg, Buzzegoli:2021wlg}:
\begin{equation}\label{eq:S_LE}
    S^\mu(p) \;=\; -\,\frac{1}{8m}\,\epsilon^{\mu\nu\rho\sigma}\,p_\nu\,\frac{\int_{\Sigma_\FO} d\Sigma\cdot p\;n_F(1-n_F)\,\varpi_{\rho\sigma}}{\int_{\Sigma_\FO} d\Sigma\cdot p\;n_F}\;+\;O(\varpi^3)\,,
\end{equation}
which is linear in $\varpi$ at leading order. The observed mean polarization $\langle P_y\rangle$ is obtained from $S^\mu(p)$ by the standard heavy-ion projection-and-integration procedure (\cite{Karpenko:2016jyx}, Eq.~(8) and Eqs.~(10)--(12)): the polarization four-vector $P^\mu(p) = 2 S^\mu(p)$ (with $s=1/2$ for the $\Lambda$) is Lorentz-boosted to the $\Lambda$ rest frame, momentum-integrated over the produced-$\Lambda$ distribution weighted by the Cooper--Frye flux $d\Sigma_\lambda p^\lambda f(x,p)$, and projected onto the global-orbital-angular-momentum direction $\hat y$, the latter being the coordinate convention of~\cite[Sec.~4.1]{Karpenko:2016jyx}. Because the underlying $S^\mu(p)$ in~\eqref{eq:S_LE} is linear in $\varpi$ at leading order, the resulting $\langle P_y\rangle$ inherits the same linearity, $\langle P_y\rangle = a\,\bar\varpi/T + O(\varpi^3)$, with $\bar\varpi$ the $\varpi^{\mu\nu}$ value averaged under this Cooper--Frye-flux-weighted measure and $a$ a dimensionless coefficient absorbing the geometric and decay factors. On the strange-quark side, applying the same Becattini-type LE spin-polarization formula~\eqref{eq:S_LE} at the constituent-quark level, i.e.\ to thermal strange quarks in the QGP phase at the particlization hypersurface, gives an analogous expression $P_s = b\,\bar\varpi/T + O(\varpi^3)$~\footnote{The thermal-vorticity scaling of constituent-quark polarization at local equilibrium is not the form of $P_q$ originally used in the coalescence literature: the authors of~\cite{Liang:2004xn} derived $P_q$ from microscopic polarized parton-parton scattering in an effective static potential model (their Eq.~(1), with $P_q \propto \mu p/[E(E+m_q)]$ and $\mu$ the Debye screening mass), which is a microscopic spin-orbit input rather than a thermal-LE result. The LE-vorticity form $P_q\propto \varpi/T$ used here is the same formula~\eqref{eq:S_LE} evaluated at the constituent-quark level. The two pictures are complementary inputs to the coalescence formulas~\cite{Liang:2004xn,Yang:2017sdk}; we adopt the LE-vorticity picture because it is the one consistent with the freeze-out chain of section~\ref{sec:setup}.}. This $P_s$ is then the polarization input to the coalescence formulas of~\cite{Liang:2004xn,Yang:2017sdk}. Eliminating $\bar\varpi$ between the two integrated forms gives $\langle P_y\rangle = (a/b)\,P_s$, i.e.\ $\langle P_y\rangle\propto P_q$ to leading order, with $q=s$ for the $\Lambda$, but the transfer factor $a/b$ is model dependent~\footnote{The common order-unity estimate is a naive coalescence expectation, not a model-independent theorem. In the $SU(6)$ quark model the $\Lambda$ spin is carried entirely by the strange quark, the $ud$ pair sitting in an isosinglet spin-$0$ antisymmetric state. The corresponding relation $P_\Lambda = P_s$ between the $\Lambda$ polarization and the strange-quark polarization is stated explicitly in~\cite{Liang:2004ph}. In this toy model the observable $\langle P_y\rangle$ along $\hat y$ is, for the directly-coalesced exclusive recombination contribution, equal to the strange-quark polarization $P_s$ along $\hat y$, so $a/b\simeq1$ and the coefficient $C$ in~\eqref{eq:naive_curve} approaches $-4/9$. Realistic deviations from this naive identification come from the mass-dependent kinematic factors in~\eqref{eq:S_LE} when applied at $m=m_\Lambda$ versus $m=m_s$, acceptance and momentum weighting in the Cooper--Frye average, and feed-down corrections from heavier hyperon decays $\Sigma^0,\Sigma(1385),\Lambda(1405),\ldots\to\Lambda+X$ (which alter the proportionality between primary-$\Lambda$ polarization and the measured $\langle P_y\rangle$ by roughly $15\%$, as computed in~\cite{Karpenko:2016jyx}, Eq.~(15)). We therefore keep this transfer factor explicit below rather than treating its numerical value as fixed.}. Combining with~\eqref{eq:rho00_coalescence} under $P_{\bar q}\approx P_q$ then produces the parametric relation
\begin{equation}\label{eq:naive_curve}
    \rho_{00}^\phi - \tfrac{1}{3} \;\approx\; -\,\tfrac{4}{9}\,r^2\,\langle P_y\rangle^2 \;\equiv\; C\,\langle P_y\rangle^2\,,
\end{equation}
with $r\equiv P_q/\langle P_y\rangle=b/a$ the quark-to-$\Lambda$ polarization-transfer ratio fixed by the coalescence and freeze-out kernels. In the naive $SU(6)$ direct-coalescence estimate $r\simeq1$, so $C$ is a negative coefficient of order unity; outside that estimate, $r$ should be regarded as a model-dependent input.

The smallness of this quadratic prediction relative to data, and in particular its predicted sign opposite to what is observed, is the original motivation for adding the local field-correlation contribution proposed in~\cite{Sheng:2019kmk, Sheng:2022wsy}, which we denote below by the SLW response sector. This relation traces out a parabola in the $(\langle P_y\rangle,\rho_{00}^\phi-1/3)$ plane, and is precisely the image $\mathrm{Im}(\widetilde{\mathfrak{F}}_{\Lambda}\times\widetilde{\mathfrak{F}}_{\phi}|_{S})$ of the one-dimensional accessible slice $S$ in the language of the Operational test of section~\ref{sec:consistency}. Riding along this parabola to the experimentally observed $\langle P_y\rangle\sim$ few $\times\,10^{-3}$ gives the predicted alignment
\begin{equation}\label{eq:naive_estimate}
    \rho_{00}^\phi - \tfrac{1}{3}\bigg|_{\rm curve} \;\sim\; 10^{-5}\text{--}10^{-4}\,,
\end{equation}
which is two to three orders of magnitude smaller than the measured value~\eqref{eq:rho00_value}. The geometric statement is therefore the following. In an energy- and centrality-matched comparison, fixing the observed $\langle P_y\rangle^{\rm exp}$ would make the curve~\eqref{eq:naive_curve} predict $\rho_{00}^\phi-1/3\sim 10^{-5}\text{--}10^{-4}$, whereas the BES $\phi$-alignment measurement reports $\rho_{00}^{\phi,{\rm exp}}-1/3\simeq 1.8\times 10^{-2}$. The available measurements should therefore be read as an order-of-magnitude tension with the vorticity-only parabola, not as a completed same-bin test of~\eqref{eq:Graph}. A strict fibration consistency test requires the two coordinates $(\langle P_y\rangle,\rho_{00}^{\phi})$ to be evaluated in the same $\sqrt{s_{NN}}$, centrality, rapidity, transverse-momentum, and event-plane bin, or in a simulation setup that explicitly maps both observables to the same freeze-out state. Several theoretical mechanisms have been proposed to absorb the discrepancy suggested by the existing data~\cite{Yang:2017sdk, Sheng:2022wsy, Sheng:2025cjk}, including local strong-force-field fluctuations coupling to $s\bar s$ pairs and modifications of the spin-density-matrix construction beyond the vorticity contribution. From the present perspective, these proposals modify either the RSH input $(\beta,\Omega,\xi)$ or the fiber observable map $\widetilde{\mathfrak{F}}_\phi$ itself; in either case, the cross-observable consistency relation~\eqref{eq:Graph} is the precise statement that any candidate explanation must respect when simultaneously confronting both $\Lambda$ polarization and $\phi$ alignment data in the same kinematic bin. A diagnostic of partial pseudo-gauge mismatch would be a deviation from~\eqref{eq:Graph} that is reduced by switching the pseudo-gauge representative used for $\widetilde{\mathfrak{F}}_\phi$ while keeping that of $\widetilde{\mathfrak{F}}_\Lambda$ fixed, indicating that the two observables have been computed in implicitly inequivalent representations of the fiber.

\paragraph{Narrow versus extended freeze-out data.} This interpretation is particularly important for the local field-correlation mechanism proposed in~\cite{Sheng:2022wsy}. That result does not contradict the consistency statement above. Rather, it indicates that the narrow freeze-out data space implicit in the vorticity-dominated closure,
\begin{equation}
    R_{\FO} \sim \bigl(T^{\mu\nu},\,S^{\lambda,ab},\,j^\mu;\,\beta,\Omega,\xi\bigr)\big|_{\Sigma_\FO}\,,
\end{equation}
is too small for the $\phi$-meson spin-density matrix. This mechanism introduces additional local response data through a modified constituent-quark polarization input to the vector-meson coalescence kernel: concretely, in~\cite[Eq.~(7)]{Sheng:2022wsy} the strange-quark polarization four-vector acquires a contribution from the $\phi$-meson field-strength tensor on top of the thermal vorticity,
\begin{equation}\label{eq:SLW_quark_pol}
    P^\mu_{s}(x,\mathbf{p}) \;\approx\; \frac{1}{4m_s}\,\epsilon^{\mu\nu\rho\sigma}\left(\omega_{\rho\sigma} + \frac{g_\phi}{(u\cdot p)T_{\rm h}}F^\phi_{\rho\sigma}\right)p_\nu\,,
\end{equation}
with $g_\phi$ the effective $s\bar s\phi$ coupling and $T_{\rm h}$ the hadronization temperature. This modified $P^\mu_s$ propagates through the standard coalescence kernel of~\cite[Eq.~(4)]{Sheng:2022wsy} to produce~\cite[Eq.~(8)]{Sheng:2022wsy}, in which the resulting $\rho_{00}$ acquires explicit dependence on local $\phi$-field electric and magnetic correlations $\langle (g_\phi\mathbf{E}^\phi_i/T_{\rm h})^2\rangle$ and $\langle (g_\phi\mathbf{B}^\phi_i/T_{\rm h})^2\rangle$ beyond the standard vorticity terms; complementary realizations are discussed in~\cite{Sheng:2019kmk,Sheng:2022ffb,Chen:2023hqi,Kumar:2023ghs}. In the present language this amounts to replacing the minimal freeze-out data space by an extended one,
\begin{equation}\label{eq:RFO_ext}
    R_{\FO}^{\rm ext}
    =
    R_{\FO}\times \mathcal{R}_{\rm SLW}\,,
\end{equation}
and, if the added responses are thermodynamically constrained at freeze-out, replacing the generator by
\begin{equation}\label{eq:KLE_ext}
    \hat K_{\LE}^{\rm ext}
    =
    \hat K_{\LE}
    +
    \int_{\Sigma_\FO} d\Sigma_\mu\,
    \sum_A \lambda_A(x)\,\hat{\mathcal R}_A^\mu(x)\,,
    \qquad
    \hat{\mathcal R}_A^\mu\in\mathcal{R}_{\rm SLW}\,.
\end{equation}
For the specific SLW realization, the additional responses can be identified concretely as composite $\phi$-field operators. Reading~\eqref{eq:SLW_quark_pol} and the resulting~\cite[Eq.~(8)]{Sheng:2022wsy}, the field-dependent part of $\rho_{00}^\phi$ at a hadronization point $x\in\Sigma_\FO$ depends on the local two-point $\phi$-field correlator $\langle :F^\phi_{\mu\nu}(x)F^\phi_{\alpha\beta}(x):\rangle$, projected by SLW onto the anisotropic parameters $F^2_T=\langle(g_\phi\mathbf{E}^\phi_{x,y}/T_{\rm h})^2\rangle$ and $F^2_z=\langle(g_\phi\mathbf{E}^\phi_z/T_{\rm h})^2\rangle$ (with corresponding magnetic components) fitted to STAR data in their Fig.~1(b). If one wishes to reproduce this functional dependence from a maximum-entropy construction, $\rho_{00}^\phi = \mathrm{Tr}(\hat\rho_{\LE}^{\rm ext}\,\hat\rho_{00\,{\rm op}}^\phi)$, a natural effective ansatz is to include the composite operator $:F^\phi_{\mu\nu}F^\phi_{\alpha\beta}:(x)$ in the constrained operator set, since every additional expectation value constrained on $\Sigma_\FO$ generates exactly one Lagrange multiplier in $\hat K_\LE$ by Legendre transform. This inclusion is not uniquely mandated by maximum entropy---such a correlator may instead enter through a coalescence model or hadronization kernel without being an independent thermodynamic constraint---but it provides a systematic way to incorporate the SLW mechanism into the LE framework. The corresponding instantiation of~\eqref{eq:KLE_ext} therefore takes the concrete form
\begin{equation}\label{eq:KLE_SLW}
    \hat K_{\LE}^{\rm ext}
    \;\supset\;
    \int_{\Sigma_\FO} d\Sigma_\mu\;
    \Lambda^{\mu\nu\rho,\alpha\beta}(x)\,
    :\!F^\phi_{\nu\rho}(x)\,F^\phi_{\alpha\beta}(x)\!:\,,
\end{equation}
with $\Lambda^{\mu\nu\rho,\alpha\beta}(x)$ the Lagrange multiplier conjugate to the local $\phi$-field-fluctuation tensor on $\Sigma_\FO$. Its independent anisotropic projections are precisely SLW's $F^2_T$ and $F^2_z$, so $\mathcal{R}_{\rm SLW}$ in~\eqref{eq:RFO_ext} is concretely populated by the composite $:F^\phi F^\phi:(x)$, with~\eqref{eq:KLE_SLW} the corresponding extended-LE generator term.\footnote{This is an extended-LE reinterpretation of the SLW mechanism, not a literal derivation in~\cite{Sheng:2022wsy}, which is formulated through the vector-meson spin Boltzmann equation and coalescence kernel. The local composite $:F^\phi F^\phi:(x)$ should be understood as the hadron-size-smeared coincident limit of the short-distance field correlator, consistent with SLW's approximation that variations of the fields inside the hadron size can be neglected.}

It is important to address a possible objection at this point: adding a new operator to $\hat K_\LE$ is necessary but not by itself sufficient to lift the dimension of the experimentally accessible slice $S^{\rm ext}$ above one. If the new Lagrange multiplier $\Lambda^{\mu\nu\rho,\alpha\beta}(x)$ in~\eqref{eq:KLE_SLW} happened to be a fixed function of the thermal multipliers, $\Lambda = g(\beta,\Omega,\xi)$, varying it would merely deform the parabolic curve~\eqref{eq:naive_curve} into a different 1-d locus rather than producing a 2-d region of $\mathbb{R}^2$. The increase $\dim S^{\rm ext}>\dim S$ requires {\em independent} variation of the SLW-sector multiplier from the thermal-sector multipliers. At the manifold level of $\mathcal{Q}_\FO^{\rm ext}$ this independence is automatic, since each constrained expectation in the maximum-entropy construction generates a logically independent Lagrange multiplier by Legendre transform; but at the level of the physically populated slice $S^{\rm ext}$ it is a substantive empirical and dynamical question.

We propose three arguments that support the independence in the SLW case. First, $\varpi$ and $\langle FF \rangle$ have different physical origins: the thermal vorticity $\varpi^{\mu\nu}$ is set by QGP relativistic hydrodynamics, namely by initial-state orbital angular momentum redistributed by viscous flow evolution, whereas $\langle :F^\phi_{\mu\nu}F^\phi_{\alpha\beta}:\rangle$ is set by non-perturbative gluon-field dynamics at hadronization, operating at the QCD scale rather than the hydrodynamic scale. There is no known dynamical mechanism that constrains one as a function of the other. Second, one can see that Sheng et al.~\cite[Fig.~1(b)]{Sheng:2022wsy} extract $F^2_T$ and $F^2_z$ as independent parameters at each collision energy, with non-trivial $\sqrt{s_{NN}}$-dependence that is not a simple function of the energy dependence of $\bar\varpi/T$ as reconstructed from $\langle P_y\rangle$. The fit is non-trivial precisely because $F^2_T, F^2_z$ are not constrained to lie on the parabola~\eqref{eq:naive_curve}. Third, we notice that the $K^{*0}$ alignment is consistent with $1/3$ in the same kinematic windows where $\rho_{00}^\phi-1/3\simeq 1.8\times 10^{-2}$~\cite{STAR:2022fan}. If the extra field-correlation response were a universal function of the vortical background, the two vector mesons measured in the same collision and kinematic windows would track each other up to quark-content and coalescence-kernel factors rather than differing by orders of magnitude; the observed $\phi$-versus-$K^{*0}$ species dependence is itself evidence that the field-correlation sector varies independently of the vortical sector. Together these arguments show that $S^{\rm ext}$ is generically a two- or higher-dimensional slice of $\mathcal{Q}_\FO^{\rm ext}$, not a deformed 1-d locus. A hidden functional dependence $\Lambda = g(\beta,\Omega,\xi)$ that would collapse $S^{\rm ext}$ back to a curve cannot be excluded {\em a priori}, but it would require fine-tuned cancellation between dynamically separate sectors and is already disfavored by the SLW fit and by the observed vector-meson species dependence.

The universal stabilizer is then the corresponding extended stabilizer
\begin{equation}\label{eq:Huniv_ext}
    H_{\univ}^{\rm ext}
    =
    \bigl\{\Phi:\Delta_\Phi\hat K_{\LE}^{\rm ext}=c_\Phi\mathbf 1\bigr\}\,,
    \qquad
    \mathcal{Q}_{\FO}^{\rm ext}
    =
    R_{\FO}^{\rm ext}/H_{\univ}^{\rm ext}\,.
\end{equation}
If the extra responses transform under a generalized pseudo-gauge operation, their variations contribute new terms to $\Delta_\Phi\hat K_{\LE}^{\rm ext}$ and hence enlarge the fiber directions probed by $\rho_{00}^\phi$. If they are PGT-inert but enter only the $\phi$ coalescence map, they still enlarge the domain of $\widetilde{\mathfrak{F}}_\phi$. Thus the prediction of the generalized framework is not that SLW must lie on the narrow image $\mathrm{Im}(\widetilde{\mathfrak{F}}_{\Lambda}\times\widetilde{\mathfrak{F}}_{\phi})$, but that it should lie on the enlarged image
\begin{equation}\label{eq:Graph_ext}
    \bigl(\langle P_y\rangle,\rho_{00}^\phi\bigr)
    \in
    \mathrm{Im}\bigl(\widetilde{\mathfrak{F}}_{\Lambda}^{\rm ext}
    \times
    \widetilde{\mathfrak{F}}_{\phi}^{\rm ext}\bigr)
    \subset
    \mathcal{M}_\mathcal{O}^{(\Lambda)}
    \times
    \mathcal{M}_\mathcal{O}^{(\phi)}\,.
\end{equation}
In this sense the observed tension between $\Lambda$ hyperon polarization and $\phi$-meson spin alignment is a discovery target for additional freeze-out responses: once the SLW sector is included, the cross-observable consistency relation should be restored in $\mathcal{Q}_{\FO}^{\rm ext}$ rather than abandoned.

\paragraph{Phenomenological stabilizer prediction.} The extended picture makes a simple phenomenological prediction. If the SLW sector is the missing response behind the large $\phi$-meson alignment, then varying the SLW data should move the $\phi$ spin-alignment observable while leaving the already-successful order-of-magnitude description of $\Lambda$ polarization approximately intact. In other words, the added response directions should lie approximately in the observable stabilizer of $\langle P_y\rangle$, but not in the observable stabilizer of $\rho_{00}^{\phi}$.

Let $c = \{c_A\}$ collectively denote the local parameters describing the added SLW sector $\mathcal{R}_{\rm SLW}$, so that a point of $\mathcal{Q}_\FO^{\rm ext}$ is $([R],c)$. We write $\Delta_c$ for the variation along the SLW fiber directions, analogous to $\Delta_\Phi$ for the standard PGT directions. The extended observable maps then decompose as
\begin{align}
    \widetilde{\mathfrak{F}}_{\Lambda}^{\rm ext}([R],c)
    &=
    \widetilde{\mathfrak{F}}_{\Lambda}([R])
    +
    \Delta_c\langle P_y\rangle([R],c)\,,
    \\
    \widetilde{\mathfrak{F}}_{\phi}^{\rm ext}([R],c)
    &=
    \widetilde{\mathfrak{F}}_{\phi}([R])
    +
    \Delta_c\rho_{00}^{\phi}([R],c)\,,
\end{align}
where $\Delta_c$ vanishes at $c=0$.
For the SLW sector to resolve the $\phi$-alignment excess without spoiling the successful order-of-magnitude description of $\Lambda$ polarization, its leading response must satisfy
\begin{equation}\label{eq:SLW_stabilizer_condition}
    \frac{\partial}{\partial c_A}
    \widetilde{\mathfrak{F}}_{\Lambda}^{\rm ext}
    \simeq 0\,,
    \qquad
    \frac{\partial}{\partial c_A}
    \widetilde{\mathfrak{F}}_{\phi}^{\rm ext}
    \neq 0
    \quad
    \text{for at least some }A\,.
\end{equation}
Equivalently, the SLW directions should lie approximately in the observable stabilizer of $\Lambda$ polarization but not in that of $\phi$ spin alignment,
\begin{equation}\label{eq:SLW_obs_stabilizers}
    \delta c\in H_{\Lambda}^{\rm obs}
    :=
    \{\delta R_{\FO}^{\rm ext}:D\widetilde{\mathfrak{F}}_{\Lambda}^{\rm ext}[\delta R_{\FO}^{\rm ext}]\simeq0\}\,,
    \qquad
    \delta c\notin H_{\phi}^{\rm obs}\,.
\end{equation}
The approximate sign is essential: this is a phenomenological stabilizer, defined up to present experimental errors and uncertainties in the freeze-out/coalescence model, not an exact operator identity like~\eqref{eq:Huniv_ext}.

A minimal local model makes the diagnostic explicit. Let $\omega$ denote the single vorticity-dominated variable that controls the narrow response slice in a fixed kinematic bin, and let $c$ denote one SLW-sector amplitude, with the multi-parameter case obtained by replacing $c$ with $c_A$. Near the vorticity-dominated point one may write
\begin{align}
    \langle P_y\rangle
    &=
    A_\Lambda\,\omega
    +
    \epsilon_\Lambda\,c
    +
    O(\omega^3,c^2,\omega c)\,,
    \label{eq:toy_Lambda_response}\\
    \rho_{00}^{\phi}-\frac13
    &=
    C_\phi\,\omega^2
    +
    B_\phi\,c
    +
    O(\omega^4,c^2,\omega c)\,.
    \label{eq:toy_phi_response}
\end{align}
At $c=0$, eliminating $\omega$ gives the vorticity-only parabola $\rho_{00}^{\phi}-1/3=(C_\phi/A_\Lambda^2)\langle P_y\rangle^2$, i.e.\ the one-dimensional curve discussed around~\eqref{eq:naive_curve}. Turning on $c$ adds the new tangent direction $(\epsilon_\Lambda,B_\phi)$ in the product observable space. The SLW explanation of the large $\phi$ alignment while preserving the successful order-of-magnitude description of $\Lambda$ polarization is therefore the statement that, after normalizing by experimental and model uncertainties, $|\epsilon_\Lambda|\ll |B_\phi|$. This is the toy-model form of the approximate stabilizer condition~\eqref{eq:SLW_stabilizer_condition}.

The physical reason such a hierarchy can occur is that the two observables probe different tensor ranks of the spin/field sector. The global $\Lambda$ polarization is a rank-one spin-vector observable, sensitive primarily to a mean spin polarization along the orbital-angular-momentum direction. By contrast, $\rho_{00}^\phi-1/3$ is a rank-two tensor-alignment observable; in particular, $\rho_{00}\neq1/3$ need not signal only a global quark polarization, but may also arise from local spin alignment or local field correlations~\cite{Xia:2020tyd,Sheng:2022wsy,Kumar:2023ghs}. A local field sector can therefore be invisible, or nearly invisible, to $\langle P_y\rangle$ while remaining visible to $\rho_{00}^\phi$: for a fluctuating field-like variable $\mathcal{B}_i$ with
\begin{equation}
    \langle \mathcal{B}_i\rangle \simeq 0\,,
    \qquad
    \mathcal{C}_{ij}:=
    \langle \mathcal{B}_i\mathcal{B}_j\rangle
    -\frac13\delta_{ij}\langle \mathcal{B}^2\rangle
    \neq 0\,,
\end{equation}
a vector polarization receives no leading contribution from the mean field, whereas a tensor spin alignment can receive a contribution of the schematic form
\begin{equation}\label{eq:rank2_SLW_response}
    \Delta_c\langle P_y\rangle
    \propto
    \hat y^i\langle \mathcal{B}_i\rangle
    \simeq 0\,,
    \qquad
    \Delta_c\rho_{00}^{\phi}
    \propto
    \hat y^i\hat y^j\mathcal{C}_{ij}
    \neq 0\,.
\end{equation}
This is the operational content of a response sector that is approximately invisible to $\Lambda$ polarization but visible to $\phi$-meson spin alignment. In the product space of measurements, variations of the SLW sector displace the theoretical point predominantly in the $\rho_{00}^{\phi}$ direction while leaving the $\Lambda$-polarization coordinate approximately fixed.

This gives several falsifiable consequences. First, after controlling for the vorticity-dominated variables that set $\langle P_y\rangle$, variations of the SLW sector should produce sizable changes in $\rho_{00}^{\phi}$ with only weak correlated changes in $\Lambda$ polarization:
\begin{equation}\label{eq:SLW_pheno_slope}
    \left|
    \frac{\partial \rho_{00}^{\phi}/\partial c_A}
         {\partial \langle P_y\rangle/\partial c_A}
    \right|
    \gg 1
    \quad
    \text{for the SLW-sensitive directions.}
\end{equation}
Second, the effect should be strongest for hadron species directly coupled to the additional local response data. The observed pattern that the $\phi$ alignment is large while the corresponding $K^{*0}$ alignment is consistent with $1/3$~\cite{STAR:2022fan} is therefore not an obstacle to the generalized framework; it is precisely the sort of $\phi$-versus-$K^{*0}$ species dependence expected if $\mathcal{R}_{\rm SLW}$ contains strange-quark or $\phi$-field correlation data rather than a universal vorticity correction. This species dependence should also be separated from late hadronic rescattering effects, which are themselves species dependent and can significantly modify $K^*$ and $\rho$ spin-alignment signals while being much smaller for the long-lived $\phi$~\cite{Shen:2021jcg}. Third, one should search for differential dependences of $\rho_{00}^{\phi}$ on transverse momentum, centrality, beam energy, and azimuthal angle relative to the event plane that are not mirrored one-for-one by $\Lambda$ polarization. This is also the natural place where the SLW prediction of non-trivial $p_T$ and azimuthal dependence, and related local-correlation/glasma-field scenarios, enter~\cite{Sheng:2022wsy,Sheng:2022ffb,Kumar:2023ghs}. A future observation of large, unavoidable SLW-induced shifts in $\Lambda$ polarization would not falsify the fibration logic, but it would falsify the simpler phenomenological claim~\eqref{eq:SLW_stabilizer_condition} that the added sector is approximately invisible to $\Lambda$ polarization.

We note that the local-polarization sign puzzle for $\Lambda$ hyperons~\cite{Becattini:2024aaa, Becattini:2021suc} can also be cast as a cross-observable consistency question between $S^\mu(p)$ at different azimuthal angles: simultaneous fiber observables in the same kinematic bin constrain a low-dimensional subvariety of the product observable space, and apparent tensions signal either a missing RSH ingredient or a genuine pseudo-gauge ambiguity.

\paragraph{Which operators transform under PGT?} At the operator level, only $\hat T^{\mu\nu}$ and $\hat S^{\lambda,\mu\nu}$ transform under PGT~\cite{Hehl:1976vr, Speranza:2020ilk}; internal-symmetry currents such as $\hat j^\mu$ are PGT-invariant by construction. At the level of expectation values, however, any operator inherits pseudo-gauge dependence through $\hat\rho_\LE$: for example, the axial-vortical-effect conductivity~\cite{Buzzegoli:2021wlg} differs among pseudo-gauge choices despite $\hat j_5^\mu$ being PGT-invariant as an operator. For SLW-type extensions, even PGT-inert operators alter the LE state through their inclusion in $\hat K_\LE^{\rm ext}$ and hence affect all fiber observables.

\subsection{Weyl-Anomaly-Induced Current as a Base Response}\label{sec:weyl_anomalybase}

As a concrete worked example of the base/fiber classification of section~\ref{sec:fibration}, we analyze the Weyl-anomaly-induced transport current derived in~\cite{Yang:2026weyl}. That result obtains a vector current in an accelerated relativistic fluid from the boundary QFT on a Rindler horizon. We show that this current is a base observable in the sense of~\eqref{eq:base_vs_fiber} and identify the criterion under which a local-equilibrium extension could promote it to a fiber observable.

The derivation of~\cite{Yang:2026weyl} is performed at global equilibrium: $\beta^\mu=u^\mu/T$ is a Killing field, the thermal vorticity is constant, and the electromagnetic background satisfies the equilibrium constraints. In their conventions the anomaly-induced contribution is
\begin{equation}\label{eq:Weyl_current}
    j_{\rm W}^\mu
    =
    -4C_{\rm W}\,F^{\mu\nu}a_\nu\,,
\end{equation}
where $C_{\rm W}$ is the Weyl-anomaly coefficient, $F_{\mu\nu}$ is the external electromagnetic field strength, and $a^\mu=u^\lambda\nabla_\lambda u^\mu$ is the fluid acceleration. At global equilibrium the velocity and temperature are determined by the Killing field $\beta^\mu=u^\mu/T$, so $a^\mu$ is a functional of the same thermodynamic source data; for example $u^\mu=\beta^\mu/\sqrt{\beta^2}$ and $T=(\beta^2)^{-1/2}$ in our mostly-minus convention. Thus the anomaly-induced current defines a map
\begin{equation}\label{eq:Weyl_base_map}
    \overline{\mathfrak F}_{\rm W}:\;
    \mathcal M_{\phys}^{A}
    \longrightarrow
    \mathcal M_{j}\,,
    \qquad
    (\beta,\Omega,\xi;A_\mu,C_{\rm W})
    \longmapsto
    -4C_{\rm W}F^{\mu\nu}a_\nu\,,
\end{equation}
where $\mathcal M_{\phys}^{A}$ denotes the thermodynamic base enlarged by the external electromagnetic source and by the anomaly coefficient. The spin potential $\Omega$ is included in the displayed domain only to match the rest of the paper; the leading Weyl-anomaly current in~\eqref{eq:Weyl_current} does not depend on it.

This immediately gives the fibration statement. Standard pseudo-gauge transformations act on the stress-spin complex, not on the electromagnetic source or on the thermodynamic multipliers:
\begin{equation}
    \delta_\Phi A_\mu=0,\qquad
    \delta_\Phi F_{\mu\nu}=0,\qquad
    \delta_\Phi\beta^\mu=\delta_\Phi\xi=\delta_\Phi\Omega_{ab}=0\,.
\end{equation}
Consequently the pullback of~\eqref{eq:Weyl_base_map} to $\mathcal Q_\FO$ is constant along the pseudo-gauge fiber,
\begin{equation}\label{eq:Weyl_vertical_zero}
    D\widetilde{\mathfrak F}_{\rm W}(V_\Phi)=0
    \qquad
    \text{for every }
    V_\Phi\in T\bigl(G_\PGT/H_\univ[\beta,\Omega,\xi]\bigr)\,.
\end{equation}
Equivalently,
\begin{equation}
    \widetilde{\mathfrak F}_{\rm W}
    =
    \overline{\mathfrak F}_{\rm W}\circ\pi_A\,,
\end{equation}
with $\pi_A:\mathcal Q_\FO^{A}\to\mathcal M_{\phys}^{A}$ the source-enlarged thermal projection. This is precisely the definition of a base observable. At the global-equilibrium point used in~\cite{Yang:2026weyl}, section~\ref{sec:GTE_saturation} gives the stronger statement $H_\univ=G_\PGT$, so the pseudo-gauge fiber itself degenerates to a point.

Relaxing global equilibrium to local equilibrium does not by itself make~\eqref{eq:Weyl_current} a fiber observable. If it is promoted only as a local constitutive term built from $(\beta,A_\mu,C_{\rm W})$, it remains a single-valued function on $\mathcal M_{\phys}^{A}$. What local equilibrium can do is open non-trivial pseudo-gauge fibers in the density operator. The full vector-current expectation value may then have the schematic form
\begin{equation}\label{eq:Weyl_LE_decomp}
    \langle \hat j^\mu\rangle_{\LE}
    =
    j_{\rm W}^\mu[\beta,A]
    +
    j_{\rm rem}^\mu[\beta,\Omega,\xi;\PG]
    +
    \cdots\,,
\end{equation}
where the first term is the anomaly-fixed base contribution and $j_{\rm rem}^\mu$ denotes possible local-equilibrium dressing. The current becomes fiber-sensitive exactly when this remainder has a non-zero vertical derivative. In the BKM language of section~\ref{sec:setup}, the criterion is
\begin{equation}\label{eq:Weyl_fiber_criterion}
    \delta_\Phi\langle\hat j^\mu\rangle_{\LE}
    =
    -\bigl\langle\hat j^\mu;\Delta_\Phi\hat K_\LE\bigr\rangle_\LE
    \neq 0
    \qquad
    \text{for some }\Phi\notin H_\univ[\beta,\Omega,\xi]\,.
\end{equation}
For the canonical-to-Belinfante direction this becomes, schematically,
\begin{equation}\label{eq:current_spin_susceptibility}
    (\Omega-\varpi)_{ab}
    \int_{\Sigma_\FO}d\Sigma_\lambda\,
    \bigl\langle
        \hat j^\mu(x);
        \hat S_{\can}^{\lambda,ab}(y)
    \bigr\rangle_\LE
    \neq 0\,,
\end{equation}
up to the geometric and boundary terms displayed in~\eqref{eq:K_PGT}. If this susceptibility vanishes, or if a Ward identity forces the current to depend only on $(\beta,A_\mu,C_{\rm W})$, the current remains base even away from global equilibrium. If it is non-zero, the correction is a fiber observable. Thus a current derived from a boundary or horizon construction need not be pseudo-gauge sensitive: the protected Weyl-anomaly coefficient and the global-equilibrium current~\eqref{eq:Weyl_current} are base data, while only a local-equilibrium extension satisfying~\eqref{eq:Weyl_fiber_criterion} would probe the pseudo-gauge fiber.

\section{Discussion}\label{sec:discussion}

We have given a geometric formulation of the pseudo-gauge dependence of Cooper--Frye observables at freeze-out. The central object is the universal stabilizer $H_\univ$ defined by the condition $\Delta_\Phi\hat K_\LE = c_\Phi\mathbf 1$, which is necessary and sufficient for any observable computed from $\hat\rho_\LE$ to be pseudo-gauge invariant. For each chosen observable $\mathcal O$, the Cooper--Frye readout $\mathfrak F_{\mathcal O}$ factors exactly through the quotient $\mathcal{Q}_\FO = R_\FO/H_\univ$, while the residual quotient is fibered over the space $\mathcal{M}_\phys$ of thermodynamic Lagrange multipliers with state-dependent fiber $G_\PGT/H_\univ[\beta,\Omega,\xi]$. The non-trivial fiber is the obstruction underlying the Buzzegoli effect; its dimension upper-bounds the number of independent PG-sensitive observables and supplies a target for future searches beyond the canonical--Belinfante comparison.

The two-class classification of observables, base observables on $\mathcal{M}_\phys$ versus fiber observables on $\mathcal{Q}_\FO$, gives a sharp criterion for which polarization moments are intrinsically pseudo-gauge robust and which carry residual pseudo-gauge dependence. The cross-observable consistency relations~\eqref{eq:Graph} provide multi-observable self-consistency tests that are absent from standard analyses and could be tested by simultaneous measurement of $\Lambda$ hyperon polarization~\cite{STAR:2017ckg, STAR:2018gyt} and $\phi$-meson spin alignment~\cite{STAR:2022fan} in the same kinematic bins. They also give a structural interpretation of the mechanism proposed in~\cite{Sheng:2022wsy}: the large $\rho_{00}^{\phi}-1/3$ signal need not contradict a narrow vorticity-based description of $\Lambda$ polarization, but rather points to additional freeze-out response data that are approximately invisible to the $\Lambda$ spin-vector observable and visible to the $\phi$-meson tensor-alignment observable.

For a fixed finite pseudo-gauge family and a chosen observable, one may still define an operational PG spread, for example
\begin{equation}\label{eq:PG_spread}
    \mathcal{V}_{\mathcal O}^{(\can,\Bel)}([R])
    :=
    \left\|
    \widetilde{\mathfrak F}_{\mathcal O}^{\can}([R])
    -
    \widetilde{\mathfrak F}_{\mathcal O}^{\Bel}([R])
    \right\|_{\mathcal M_{\mathcal O}}\,.
\end{equation}
This quantity vanishes at global equilibrium and is a useful indication that the chosen observable probes a residual fiber direction. It should not, however, be read as an equivalence criterion for equilibrium: it may vanish away from global equilibrium if the observable is insensitive to the chosen PG direction. The invariant statement is the stabilizer condition itself; quantities such as~\eqref{eq:PG_spread} are model- and observable-dependent probes of that geometry.

Several directions extend the present results. First, the full canonical-to-Belinfante computation on a curved $\Sigma_\FO$, rather than the planar idealization of section~\ref{sec:BC_test}, introduces extrinsic-curvature terms that modify the equilibrium relation~\eqref{eq:Locking}. The recent analysis of curved freeze-out surfaces and tensor spin alignment~\cite{Sheng:2025cjk} suggests these geometric corrections are quantitatively relevant for the $\rho_{00}$ observable. Second, the explicit form of the fiber $G_\PGT/H_\univ$ for general PGT generators $\Phi^{\lambda,\mu\nu}$ (including a non-trivial spin superpotential $Z^{ab,\mu\rho}$~\cite{Speranza:2020ilk}) would identify additional PG-sensitive observables beyond the canonical-Belinfante one already used by~\cite{Buzzegoli:2021wlg}. Third, pairing this framework with realistic RSH simulations~\cite{Florkowski:2017ruc, Florkowski:2018fap, Hongo:2021ona} would yield numerical predictions for family-restricted PG spreads such as~\eqref{eq:PG_spread} in actual collision kinematics and could turn the cross-observable consistency relations into quantitative experimental tests.

A further natural direction is the specialization to Bjorken-symmetric flow~\cite{Bjorken:1982qr}, where one can compare the universal stabilizer condition~\eqref{eq:Def_Huniv} with the residual pseudo-gauge condition $\partial_\lambda\Phi^{\lambda,\mu\nu}=0$ identified by~\cite{Drogosz:2024rbd} as the symmetric-to-symmetric condition. The two are logically distinct---\eqref{eq:Def_Huniv} is a surface-integral statement on $\Sigma_\FO$, whereas Drogosz's is a local spacetime constraint---and their relationship on a constant-proper-time hypersurface remains to be clarified.

\appendix

\section{Conventions and the BKM Non-degeneracy Lemma}\label{app:conventions}

We collect here a glossary of the principal symbols, our metric and notational conventions, and the technical facts about the Bogoliubov--Kubo--Mori inner product and the pseudo-gauge transformation that are used throughout the main text.

\paragraph{Glossary of symbols.}
Table~\ref{tab:glossary} collects the principal spaces, maps, and objects introduced in the paper, together with the equation in which each first appears.

\begin{longtable}{lll}
\caption{Glossary of the principal symbols used in the paper.}
\label{tab:glossary}\\
\hline
\textbf{Symbol} & \textbf{Description} & \textbf{Ref.}\\
\hline
\endfirsthead
\multicolumn{3}{l}{\small\emph{Table~\ref{tab:glossary} continued from previous page}}\\[2pt]
\hline
\textbf{Symbol} & \textbf{Description} & \textbf{Ref.}\\
\hline
\endhead
\hline
\multicolumn{3}{r}{\small\emph{Continued on next page}}\\
\endfoot
\hline
\endlastfoot
\multicolumn{3}{l}{\emph{Spaces}}\\[2pt]
$M_4$ & spacetime slab with freeze-out boundary & \S\ref{sec:intro}\\
$\Sigma_\FO$ & freeze-out hypersurface ($\subset\partial M_4$) & \S\ref{sec:intro}\\
$R_\FO$ ($R_\FO^{\LE}$) & space of (local-equilibrium) freeze-out surface data & \eqref{eq:CFmap}\\
$\mathcal{M}_\phys$ & thermodynamic moduli: $(\beta^\mu,\Omega_{ab},\xi)$ on $\Sigma_\FO$ & \eqref{eq:Mphys}\\
$\mathcal{Q}_\FO$ & quotient $R_\FO/H_\univ$; universal PGT quotient of freeze-out data & \eqref{eq:Inequiv_rho}\\
$\mathcal{M}_\mathcal{O}$ & value space of observable $\mathcal{O}$ & \eqref{eq:CFmap}\\
$\mathcal{W}$ & space of admissible Wigner functions on $\Sigma_\FO$ & \eqref{eq:Functional}\\
$V_\PGT^{(x)}$ & local PGT generator space at $x$ ($\dim=24$ in $D\!=\!4$) & \eqref{eq:Phi_dim}\\
$V_{H_\univ}^{(x)}$ & image of $H_\univ$ under the evaluation map at $x$ & \eqref{eq:V_Huniv}\\
$R_\FO^{\rm ext}$ & extended freeze-out data $R_\FO\times\mathcal{R}_{\rm SLW}$ & \eqref{eq:RFO_ext}\\[4pt]
\hline
\multicolumn{3}{l}{\emph{Groups and subgroups}}\\[2pt]
$G_\PGT$ & full pseudo-gauge transformation group & \S\ref{sec:setup}\\
$H_\univ$ & universal stabilizer: $\{\Phi\,|\,\Delta_\Phi\hat K_\LE=c_\Phi\mathbf{1}\}$ & \eqref{eq:Univ_Stab}\\
$H_\mathcal{O}$ & observable stabilizer of $\mathcal{O}$ & \eqref{eq:HO_stabilizer}\\
$F_{(\beta,\Omega,\xi)}$ & fiber $G_\PGT/H_\univ[\beta,\Omega,\xi]$ & \eqref{eq:Fiber}\\
$\mathcal{F}$ & finite family of PGT generators (span) & \eqref{eq:F_def}\\
$H_\univ^{\rm ext}$ & extended universal stabilizer for $\hat K_\LE^{\rm ext}$ & \eqref{eq:Huniv_ext}\\[4pt]
\hline
\multicolumn{3}{l}{\emph{Maps}}\\[2pt]
$\mathfrak{F}_\mathcal{O}$ & Cooper--Frye readout: $R_\FO^{\LE}\to\mathcal{M}_\mathcal{O}$ & \eqref{eq:CFmap}\\
$\widetilde{\mathfrak{F}}_\mathcal{O}$ & induced CF map on the quotient: $\mathcal{Q}_\FO\to\mathcal{M}_\mathcal{O}$ & \eqref{eq:Factorization}\\
$\overline{\mathfrak{F}}_\mathcal{O}$ & (candidate) descent to the base: $\mathcal{M}_\phys\to\mathcal{M}_\mathcal{O}$ & \eqref{eq:Diagram}\\
$\mathfrak{f}_\mathcal{O}$ & Wigner-space observable functional: $\mathcal{W}\to\mathcal{M}_\mathcal{O}$ & \eqref{eq:Functional}\\
$\mathcal{L}_\LE$ & LE assignment: $R_\FO^{\LE}\to\{\hat\rho_\LE\}$ & \eqref{eq:LE_assignment}\\
$\pi$ & thermal projection: $\mathcal{Q}_\FO\twoheadrightarrow\mathcal{M}_\phys$ & \eqref{eq:Projection_pi}\\
$q$ & quotient map: $R_\FO\twoheadrightarrow\mathcal{Q}_\FO$ & \eqref{eq:Factorization}\\
$\mathrm{ev}_x$ & evaluation map: $G_\PGT\to V_\PGT^{(x)}$ & \eqref{eq:Evaluation}\\
$\chi_W$ & BKM response: $\mathcal{A}\to\mathcal{W}$, $X\mapsto-\langle\hat W;X\rangle_\LE$ & \eqref{eq:Delta_Ophys}\\[4pt]
\hline
\multicolumn{3}{l}{\emph{Operators and fields}}\\[2pt]
$\hat K_\LE^{(\PG)}$ & local-equilibrium generator in pseudo-gauge PG & \eqref{eq:Define_K}\\
$\hat\rho_\LE^{(\PG)}$ & LE density operator $e^{-\hat K_\LE}/Z_\LE$ & \eqref{eq:Define_rho}\\
$\Phi^{\lambda,\mu\nu}$ & PGT superpotential ($\Phi^{\lambda,\mu\nu}=-\Phi^{\lambda,\nu\mu}$) & \eqref{eq:PGT}\\
$X^{\lambda\mu\nu}$ & stress-tensor shift tensor $\tfrac12(\Phi^{\lambda,\mu\nu}\!-\!\Phi^{\mu,\lambda\nu}\!-\!\Phi^{\nu,\lambda\mu})$ & \eqref{eq:PGT}\\
$\beta^\mu$ & inverse-temperature four-vector $u^\mu/T$ & \eqref{eq:Define_K}\\
$\Omega_{ab}$ & spin potential (Lagrange multiplier for $\hat S^{\mu,ab}$) & \eqref{eq:Define_K}\\
$\xi$ & thermal chemical potential $\mu/T$ & \eqref{eq:Define_K}\\
$\varpi_{\mu\nu}$ & thermal vorticity $-\partial_{[\mu}\beta_{\nu]}$ & \eqref{eq:beta_general}\\
$\xi_{\mu\nu}$ & thermal shear $\partial_{(\mu}\beta_{\nu)}$ & \S\ref{sec:BC_test}\\
$W(x,p)$ & Wigner function on $\Sigma_\FO$ & \eqref{eq:Chain}\\[4pt]
\hline
\multicolumn{3}{l}{\emph{Fiber dimensions}}\\[2pt]
$d_F(x)$ & local (pointwise) fiber dimension at $x\in\Sigma_\FO$ & \eqref{eq:dFlocal}\\
$d_F^\mathcal{F}$ & family-restricted fiber dimension for family $\mathcal{F}$ & \eqref{eq:dFfamily}\\[2pt]
\end{longtable}

\paragraph{Metric and notation.} We adopt mostly-minus metric signature $\eta_{\mu\nu}=\mathrm{diag}(+,-,-,-)$, with $\epsilon^{0123}=+1$. The unit normal $n_\mu$ to $\Sigma_\FO$ is future-directed, $n^\mu n_\mu = 1$. The surface integration measure is $d\Sigma_\mu = n_\mu\,d^3\sigma$. The Levi-Civita symbol $\epsilon_{abcd}$ is the totally antisymmetric symbol, with frame indices $a,b\in\{0,1,2,3\}$.

\paragraph{Derivation of the density-operator variation.} Let
\begin{equation}
    \hat K_\LE(\epsilon)
    =
    \hat K_\LE
    +
    \epsilon\,\Delta_\Phi\hat K_\LE
    +
    O(\epsilon^2)\,,
    \qquad
    \hat\rho_\LE(\epsilon)
    =
    \frac{e^{-\hat K_\LE(\epsilon)}}{Z_\LE(\epsilon)}\,,
    \qquad
    Z_\LE(\epsilon)=\mathrm{Tr}\,e^{-\hat K_\LE(\epsilon)}\,,
\end{equation}
where $\Delta_\Phi\hat K_\LE$ is the PGT insertion displayed in~\eqref{eq:K_PGT}. Since $\hat K_\LE$ need not commute with $\Delta_\Phi\hat K_\LE$, the first variation of the exponential is given by the Duhamel formula
\begin{equation}
    \delta e^{-\hat K_\LE}
    =
    -\int_0^1 ds\;
    e^{-(1-s)\hat K_\LE}
    \left(\Delta_\Phi\hat K_\LE\right)
    e^{-s\hat K_\LE}\,.
\end{equation}
Dividing by $Z_\LE$ and using $e^{-a\hat K_\LE}=Z_\LE^a\hat\rho_\LE^{\,a}$ gives
\begin{equation}
    \frac{1}{Z_\LE}\,\delta e^{-\hat K_\LE}
    =
    -\int_0^1 ds\;
    \hat\rho_\LE^{\,1-s}
    \left(\Delta_\Phi\hat K_\LE\right)
    \hat\rho_\LE^{\,s}\,.
\end{equation}
The variation of the partition function is, by cyclicity of the trace,
\begin{equation}
    \delta Z_\LE
    =
    \mathrm{Tr}\,\delta e^{-\hat K_\LE}
    =
    -Z_\LE\,\langle \Delta_\Phi\hat K_\LE\rangle_\LE\,,
    \qquad
    \delta\!\left(\frac{1}{Z_\LE}\right)
    =
    \frac{1}{Z_\LE}\,\langle \Delta_\Phi\hat K_\LE\rangle_\LE\,.
\end{equation}
Therefore
\begin{align}
    \Delta_\Phi\hat\rho_\LE
    &=
    \frac{1}{Z_\LE}\,\delta e^{-\hat K_\LE}
    +
    \hat\rho_\LE\,\langle \Delta_\Phi\hat K_\LE\rangle_\LE
    \nonumber\\
    &=
    -\int_0^1 ds\;
    \hat\rho_\LE^{\,1-s}
    \left(
    \Delta_\Phi\hat K_\LE
    -
    \langle\Delta_\Phi\hat K_\LE\rangle_\LE
    \right)
    \hat\rho_\LE^{\,s}\,,
\end{align}
which is~\eqref{eq:Delta_rho}. The subtraction of $\langle\Delta_\Phi\hat K_\LE\rangle_\LE$ is exactly the normalization contribution from $Z_\LE$, and ensures $\mathrm{Tr}\,\Delta_\Phi\hat\rho_\LE=0$ in order for the density operator to remain normalized.

\paragraph{The Bogoliubov--Kubo--Mori inner product.} For a density operator $\hat\rho$ on a Hilbert space $\mathcal H$ and operators $X,Y$, the normalized BKM inner product is
\begin{equation}\label{eq:BKM_def}
    (X,Y)_\rho \;:=\; \int_0^1 ds \;\mathrm{Tr}\!\left(\rho^{1-s}X^\dagger \rho^{s}Y\right)\,.
\end{equation}
On the algebra of bounded operators on a finite-dimensional $\mathcal H$, this defines a positive-definite sesquilinear form~\cite{petz1993bogoliubov, petz2007quantum, amari2000methods}. The connected BKM correlator with respect to the local-equilibrium state $\hat\rho_\LE = e^{-\hat K_\LE}/Z_\LE$ that we use throughout the main text is
\begin{equation}\label{eq:BKM_connected}
    \langle X; Y\rangle_\LE \;:=\; (X,Y)_{\hat\rho_\LE}\,-\,\langle X\rangle_\LE\langle Y\rangle_\LE \;=\; \int_0^1 ds\;\mathrm{Tr}\!\left(\hat\rho_\LE^{\,1-s}X^\dagger\hat\rho_\LE^{\,s}Y\right)\,-\,\langle X\rangle_\LE\langle Y\rangle_\LE\,.
\end{equation}
A direct computation, using $\mathrm{Tr}(\hat\rho_\LE^{1-s}\mathbf{1}\hat\rho_\LE^s Y)=\langle Y\rangle_\LE$, shows that the connected form equals the uncentered form evaluated on $\hat\rho_\LE$-centered operators:
\begin{equation}\label{eq:BKM_centered}
    \langle X; Y\rangle_\LE \;=\; (\tilde X,\tilde Y)_{\hat\rho_\LE}\,, \qquad \tilde X := X - \langle X\rangle_\LE \mathbf{1}\,,\qquad \tilde Y := Y - \langle Y\rangle_\LE \mathbf{1}\,.
\end{equation}

\paragraph{Non-degeneracy on hermitian operators modulo $\mathbb{C}\mathbf{1}$.} The essential input to the necessity direction of~\eqref{eq:Huniv_iff} is the statement: {\em if $Y$ is hermitian and $\langle X; Y\rangle_\LE = 0$ for every hermitian $X$, then $Y \in \mathbb{C}\,\mathbf 1$.} The hermiticity assumption on $Y$ is automatically satisfied in our application, since $Y=\Delta_\Phi\hat K_\LE$ is the variation of the hermitian operator $\hat K_\LE$ under a real-valued PGT parameter $\Phi$, and is therefore itself hermitian. The proof below is formulated in a UV-regulated finite-volume setting where operators are bounded and traces are finite; the result is then used formally in the continuum limit.

We give the proof in two steps. First, the BKM form $(X,X)_{\hat\rho_\LE}$ is non-negative on every operator $X$:
\begin{equation}\label{eq:BKM_positivity}
    (X,X)_{\hat\rho_\LE} \;=\; \int_0^1 ds\;\mathrm{Tr}\!\left(\hat\rho_\LE^{\,1-s}\,X^\dagger \hat\rho_\LE^{\,s}\,X\right) \;=\; \int_0^1 ds\;\bigl\|\,\hat\rho_\LE^{s/2}\,X\,\hat\rho_\LE^{(1-s)/2}\,\bigr\|_{\rm HS}^2 \;\geq\; 0\,,
\end{equation}
where $\|\cdot\|_{\rm HS}$ denotes the Hilbert--Schmidt norm. Equality holds if and only if the integrand vanishes for almost all $s$, hence for all $s$ by continuity, which for invertible $\hat\rho_\LE$ forces $X=0$. The form $(\cdot,\cdot)_{\hat\rho_\LE}$ is therefore strictly positive-definite.

Second, by~\eqref{eq:BKM_centered}, $\langle X;Y\rangle_\LE = (\tilde X,\tilde Y)_{\hat\rho_\LE}$ where $\tilde X = X - \langle X\rangle_\LE\mathbf{1}$ and $\tilde Y = Y - \langle Y\rangle_\LE\mathbf{1}$ are the $\hat\rho_\LE$-centered operators and both are hermitian when $X$ and $Y$ are. We are given that $\langle X;Y\rangle_\LE = 0$ for every hermitian $X$, which is equivalent to
\begin{equation}
    (\tilde X,\,\tilde Y)_{\hat\rho_\LE} \;=\; 0 \qquad \text{for every hermitian } \tilde X \text{ with } \langle\tilde X\rangle_\LE = 0\,.
\end{equation}
Since $\tilde Y$ itself satisfies $\langle\tilde Y\rangle_\LE = \langle Y\rangle_\LE - \langle Y\rangle_\LE = 0$, the operator $\tilde Y$ lies in the same subspace over which $\tilde X$ is taken. We are therefore free to choose $\tilde X = \tilde Y$, giving
\begin{equation}
    (\tilde Y,\,\tilde Y)_{\hat\rho_\LE} \;=\; 0\,.
\end{equation}
By positive-definiteness, $\tilde Y = 0$, hence $Y = \langle Y\rangle_\LE\,\mathbf 1\in\mathbb{C}\,\mathbf 1$. This is the precise non-degeneracy statement invoked in the proof of the universal stabilizer theorem.

\paragraph{Pseudo-gauge transformation conventions.} For completeness we record the standard pseudo-gauge transformation conventions following~\cite{Hehl:1976vr, Becattini:2018duy, Speranza:2020ilk, Drogosz:2024rbd}. The most general PGT is parameterized by a rank-3 generator $\Phi^{\lambda,\mu\nu}$ antisymmetric in its last pair, $\Phi^{\lambda,\mu\nu} = -\Phi^{\lambda,\nu\mu}$, together with an optional rank-4 superpotential $Z^{ab,\mu\rho} = -Z^{ba,\mu\rho} = -Z^{ab,\rho\mu}$. The transformation rules on the stress-energy and spin tensors are
\begin{align}
    T^{\prime\mu\nu} &\;=\; T^{\mu\nu} + \partial_\lambda X^{\lambda\mu\nu}\,, \qquad X^{\lambda\mu\nu} = \tfrac12\bigl(\Phi^{\lambda,\mu\nu}-\Phi^{\mu,\lambda\nu}-\Phi^{\nu,\lambda\mu}\bigr)\,, \\
    S^{\prime\lambda,\mu\nu} &\;=\; S^{\lambda,\mu\nu} - \Phi^{\lambda,\mu\nu} + \partial_\rho Z^{\mu\nu,\lambda\rho}\,.
\end{align}
The combined transformation preserves the total angular momentum current $J^{\lambda,\mu\nu} = x^\mu T^{\lambda\nu} - x^\nu T^{\lambda\mu} + S^{\lambda,\mu\nu}$ up to total derivatives, and the conserved baryon current $j^\mu$ is invariant. In the planar $\Sigma_\FO\cong\mathbb{R}^3$ idealization used in the main text, the $Z$-superpotential term vanishes by antisymmetry of $Z^{ab,\mu\rho}$ in its last pair and the surface integration measure $d\Sigma_\mu = n_\mu d^3\sigma$ on a constant-time surface.

\acknowledgments

JT would like to thank Professor Zuo-tang Liang for his talk on various aspects of the physics of quark-gluon plasma and its relation to relativistic spin hydrodynamics. This work is supported by National Natural Science Foundation of China under Grant No. 12405085 and by the Natural Science Foundation of Shanghai (Grant No. 24ZR1419300).



\bibliographystyle{JHEP}
\bibliography{biblio.bib}

@article{Huovinen:2012is,
    author = "Huovinen, Pasi and Petersen, Hannah",
    title = "{Particlization in hybrid models}",
    eprint = "1206.3371",
    archivePrefix = "arXiv",
    primaryClass = "nucl-th",
    doi = "10.1140/epja/i2012-12171-9",
    journal = "Eur. Phys. J. A",
    volume = "48",
    pages = "171",
    year = "2012"
}

@article{Becattini:2020ngo,
    author = "Becattini, Francesco and Lisa, Michael A.",
    title = "{Polarization and Vorticity in the Quark{\textendash}Gluon Plasma}",
    eprint = "2003.03640",
    archivePrefix = "arXiv",
    primaryClass = "nucl-ex",
    doi = "10.1146/annurev-nucl-021920-095245",
    journal = "Ann. Rev. Nucl. Part. Sci.",
    volume = "70",
    pages = "395--423",
    year = "2020"
}

@article{Cooper:1974mv,
    author = "Cooper, Fred and Frye, Graham",
    title = "{Comment on the Single Particle Distribution in the Hydrodynamic and Statistical Thermodynamic Models of Multiparticle Production}",
    reportNumber = "Print-74-0742 (YESHIVA)",
    doi = "10.1103/PhysRevD.10.186",
    journal = "Phys. Rev. D",
    volume = "10",
    pages = "186",
    year = "1974"
}

@article{Hehl:1976vr,
    author = "Hehl, F. W.",
    title = "{On the Energy Tensor of Spinning Massive Matter in Classical Field Theory and General Relativity}",
    doi = "10.1016/0034-4877(76)90016-1",
    journal = "Rept. Math. Phys.",
    volume = "9",
    pages = "55--82",
    year = "1976"
}

@article{Banerjee:2012iz,
    author = "Banerjee, Nabamita and Bhattacharya, Jyotirmoy and Bhattacharyya, Sayantani and Jain, Sachin and Minwalla, Shiraz and Sharma, Tarun",
    title = "{Constraints on Fluid Dynamics from Equilibrium Partition Functions}",
    eprint = "1203.3544",
    archivePrefix = "arXiv",
    primaryClass = "hep-th",
    reportNumber = "TFR-TH-12-05, IPMU12-0037",
    doi = "10.1007/JHEP09(2012)046",
    journal = "JHEP",
    volume = "09",
    pages = "046",
    year = "2012"
}

@article{Jensen:2012jh,
    author = "Jensen, Kristan and Kaminski, Matthias and Kovtun, Pavel and Meyer, Rene and Ritz, Adam and Yarom, Amos",
    title = "{Towards hydrodynamics without an entropy current}",
    eprint = "1203.3556",
    archivePrefix = "arXiv",
    primaryClass = "hep-th",
    reportNumber = "CCTP-2012-03",
    doi = "10.1103/PhysRevLett.109.101601",
    journal = "Phys. Rev. Lett.",
    volume = "109",
    pages = "101601",
    year = "2012"
}

@article{Gallegos:2021bzp,
    author = {Gallegos, A. D. and G{\"u}rsoy, U. and Yarom, A.},
    title = "{Hydrodynamics of spin currents}",
    eprint = "2101.04759",
    archivePrefix = "arXiv",
    primaryClass = "hep-th",
    doi = "10.21468/SciPostPhys.11.2.041",
    journal = "SciPost Phys.",
    volume = "11",
    pages = "041",
    year = "2021"
}

@article{Gallegos:2022jow,
    author = "Gallegos, A. D. and Gursoy, U. and Yarom, A.",
    title = "{Hydrodynamics, spin currents and torsion}",
    eprint = "2203.05044",
    archivePrefix = "arXiv",
    primaryClass = "hep-th",
    doi = "10.1007/JHEP05(2023)139",
    journal = "JHEP",
    volume = "05",
    pages = "139",
    year = "2023"
}

@article{Hongo:2021ona,
    author = "Hongo, Masaru and Huang, Xu-Guang and Kaminski, Matthias and Stephanov, Mikhail and Yee, Ho-Ung",
    title = "{Relativistic spin hydrodynamics with torsion and linear response theory for spin relaxation}",
    eprint = "2107.14231",
    archivePrefix = "arXiv",
    primaryClass = "hep-th",
    reportNumber = "RIKEN-iTHEMS-Report-21",
    doi = "10.1007/JHEP11(2021)150",
    journal = "JHEP",
    volume = "11",
    pages = "150",
    year = "2021"
}

@article{Becattini:2014yxa,
    author = "Becattini, F. and Bucciantini, L. and Grossi, E. and Tinti, L.",
    title = "{Local thermodynamical equilibrium and the beta frame for a quantum relativistic fluid}",
    eprint = "1403.6265",
    archivePrefix = "arXiv",
    primaryClass = "hep-th",
    doi = "10.1140/epjc/s10052-015-3384-y",
    journal = "Eur. Phys. J. C",
    volume = "75",
    number = "5",
    pages = "191",
    year = "2015"
}

@article{Becattini:2018duy,
    author = "Becattini, F. and Florkowski, Wojciech and Speranza, Enrico",
    title = "{Spin tensor and its role in non-equilibrium thermodynamics}",
    eprint = "1807.10994",
    archivePrefix = "arXiv",
    primaryClass = "hep-th",
    doi = "10.1016/j.physletb.2018.12.016",
    journal = "Phys. Lett. B",
    volume = "789",
    pages = "419--425",
    year = "2019"
}

@article{Bjorken:1982qr,
    author = "Bjorken, J. D.",
    title = "{Highly Relativistic Nucleus-Nucleus Collisions: The Central Rapidity Region}",
    doi = "10.1103/PhysRevD.27.140",
    journal = "Phys. Rev. D",
    volume = "27",
    pages = "140--151",
    year = "1983"
}

@article{Drogosz:2024rbd,
    author = "Drogosz, Zbigniew and Florkowski, Wojciech and Hontarenko, Mykhailo and Ryblewski, Radoslaw",
    title = "{Dynamical constraints on pseudo-gauge transformations}",
    eprint = "2411.06249",
    archivePrefix = "arXiv",
    primaryClass = "hep-ph",
    doi = "10.1016/j.physletb.2025.139244",
    journal = "Phys. Lett. B",
    volume = "861",
    pages = "139244",
    year = "2025"
}

@article{Speranza:2020ilk,
    author = "Speranza, Enrico and Weickgenannt, Nora",
    title = "{Spin tensor and pseudo-gauges: from nuclear collisions to gravitational physics}",
    eprint = "2007.00138",
    archivePrefix = "arXiv",
    primaryClass = "nucl-th",
    doi = "10.1140/epja/s10050-021-00455-2",
    journal = "Eur. Phys. J. A",
    volume = "57",
    number = "5",
    pages = "155",
    year = "2021"
}

@article{Buzzegoli:2021wlg,
    author = "Buzzegoli, M.",
    title = "{Pseudogauge dependence of the spin polarization and of the axial vortical effect}",
    eprint = "2109.12084",
    archivePrefix = "arXiv",
    primaryClass = "nucl-th",
    doi = "10.1103/PhysRevC.105.044907",
    journal = "Phys. Rev. C",
    volume = "105",
    number = "4",
    pages = "044907",
    year = "2022"
}

@article{Liu:2021nyg,
    author = "Liu, Yu-Chen and Huang, Xu-Guang",
    title = "{Spin polarization formula for Dirac fermions at local equilibrium}",
    eprint = "2109.15301",
    archivePrefix = "arXiv",
    primaryClass = "nucl-th",
    doi = "10.1007/s11433-022-1903-8",
    journal = "Sci. China Phys. Mech. Astron.",
    volume = "65",
    number = "7",
    pages = "272011",
    year = "2022"
}

@article{Karpenko:2016jyx,
    author = "Karpenko, I. and Becattini, F.",
    title = "{Study of $\Lambda$ polarization in relativistic nuclear collisions at $\sqrt{s_\mathrm{NN}}=7.7$--200 GeV}",
    eprint = "1610.04717",
    archivePrefix = "arXiv",
    primaryClass = "nucl-th",
    doi = "10.1140/epjc/s10052-017-4765-1",
    journal = "Eur. Phys. J. C",
    volume = "77",
    number = "4",
    pages = "213",
    year = "2017"
}

@article{Liang:2004ph,
    author = "Liang, Zuo-Tang and Wang, Xin-Nian",
    title = "{Globally polarized quark-gluon plasma in non-central A+A collisions}",
    eprint = "nucl-th/0410079",
    archivePrefix = "arXiv",
    reportNumber = "LBNL-56383",
    doi = "10.1103/PhysRevLett.94.102301",
    journal = "Phys. Rev. Lett.",
    volume = "94",
    pages = "102301",
    year = "2005",
    note = "[Erratum: Phys.Rev.Lett. 96, 039901 (2006)]"
}

@article{Huang:2024ffg,
    author = "Huang, Xu-Guang",
    title = "{An introduction to relativistic spin hydrodynamics}",
    eprint = "2411.11753",
    archivePrefix = "arXiv",
    primaryClass = "nucl-th",
    doi = "10.1007/s41365-025-01784-3",
    journal = "Nucl. Sci. Tech.",
    volume = "36",
    number = "11",
    pages = "208",
    year = "2025"
}

@book{amari2000methods,
  title={Methods of information geometry},
  author={Amari, Shun-ichi and Nagaoka, Hiroshi},
  volume={191},
  year={2000},
  publisher={American Mathematical Soc.}
}

@article{Kubo:1957mj,
    author = "Kubo, Ryogo",
    title = "{Statistical-Mechanical Theory of Irreversible Processes. I. General Theory and Simple Applications to Magnetic and Conduction Problems}",
    journal = "J. Phys. Soc. Jap.",
    volume = "12",
    number = "6",
    pages = "570--586",
    year = "1957",
    doi = "10.1143/JPSJ.12.570"
}

@book{Kubo:1991stat,
    author = "Kubo, Ryogo and Toda, Morikazu and Hashitsume, Natsuki",
    title = "{Statistical Physics II: Nonequilibrium Statistical Mechanics}",
    series = "Springer Series in Solid-State Sciences",
    volume = "31",
    publisher = "Springer",
    address = "Berlin, Heidelberg",
    edition = "2",
    year = "1991",
    doi = "10.1007/978-3-642-58244-8"
}

@article{Sheng:2025cjk,
    author = "Sheng, Xin-Li and Becattini, Francesco and Roselli, Daniele",
    title = "{An improved formula for Wigner function and spin polarization in a decoupling relativistic fluid at local thermodynamic equilibrium}",
    eprint = "2509.14301",
    archivePrefix = "arXiv",
    primaryClass = "nucl-th",
    month = "9",
    year = "2025"
}

@article{STAR:2017ckg,
    author = "Adamczyk, L. and others",
    collaboration = "STAR",
    title = "{Global $\Lambda$ hyperon polarization in nuclear collisions: evidence for the most vortical fluid}",
    eprint = "1701.06657",
    archivePrefix = "arXiv",
    primaryClass = "nucl-ex",
    doi = "10.1038/nature23004",
    journal = "Nature",
    volume = "548",
    pages = "62--65",
    year = "2017"
}

@article{STAR:2018gyt,
    author = "Adam, Jaroslav and others",
    collaboration = "STAR",
    title = "{Global polarization of $\Lambda$ hyperons in Au+Au collisions at $\sqrt{s_{NN}}$ = 200 GeV}",
    eprint = "1805.04400",
    archivePrefix = "arXiv",
    primaryClass = "nucl-ex",
    doi = "10.1103/PhysRevC.98.014910",
    journal = "Phys. Rev. C",
    volume = "98",
    pages = "014910",
    year = "2018"
}

@article{STAR:2022fan,
    author = "Abdallah, M. S. and others",
    collaboration = "STAR",
    title = "{Pattern of global spin alignment of $\phi$ and $K^{*0}$ mesons in heavy-ion collisions}",
    eprint = "2204.02302",
    archivePrefix = "arXiv",
    primaryClass = "nucl-ex",
    doi = "10.1038/s41586-022-05557-5",
    journal = "Nature",
    volume = "614",
    number = "7947",
    pages = "244--248",
    year = "2023"
}

@article{Schilling:1969um,
    author = "Schilling, K. and Seyboth, P. and Wolf, G. E.",
    title = "{On the Analysis of Vector-Meson Production by Polarized Photons}",
    doi = "10.1016/0550-3213(70)90070-2",
    journal = "Nucl. Phys. B",
    volume = "15",
    pages = "397--412",
    year = "1970"
}

@article{Becattini:2011zlx,
    author = "Becattini, F. and Tinti, L.",
    title = "{Thermodynamical inequivalence of quantum stress-energy and spin tensors}",
    eprint = "1101.5251",
    archivePrefix = "arXiv",
    primaryClass = "hep-th",
    doi = "10.1103/PhysRevD.84.025013",
    journal = "Phys. Rev. D",
    volume = "84",
    pages = "025013",
    year = "2011"
}

@article{Becattini:2012pp,
    author = "Becattini, F. and Tinti, L.",
    title = "{Nonequilibrium Thermodynamical Inequivalence of Quantum Stress-energy and Spin Tensors}",
    eprint = "1209.6212",
    archivePrefix = "arXiv",
    primaryClass = "hep-th",
    doi = "10.1103/PhysRevD.87.025029",
    journal = "Phys. Rev. D",
    volume = "87",
    pages = "025029",
    year = "2013"
}

@article{Florkowski:2017ruc,
    author = "Florkowski, Wojciech and Friman, Bengt and Jaiswal, Amaresh and Speranza, Enrico",
    title = "{Relativistic fluid dynamics with spin}",
    eprint = "1705.00587",
    archivePrefix = "arXiv",
    primaryClass = "nucl-th",
    doi = "10.1103/PhysRevC.97.041901",
    journal = "Phys. Rev. C",
    volume = "97",
    number = "4",
    pages = "041901",
    year = "2018"
}

@article{Becattini:2013fla,
    author = "Becattini, F. and Chandra, V. and Del Zanna, L. and Grossi, E.",
    title = "{Relativistic distribution function for particles with spin at local thermodynamical equilibrium}",
    eprint = "1303.3431",
    archivePrefix = "arXiv",
    primaryClass = "nucl-th",
    doi = "10.1016/j.aop.2013.07.004",
    journal = "Annals Phys.",
    volume = "338",
    pages = "32--49",
    year = "2013"
}

@article{Becattini:2021suc,
    author = "Becattini, F. and Buzzegoli, M. and Palermo, A.",
    title = "{Spin-thermal shear coupling in a relativistic fluid}",
    eprint = "2103.10917",
    archivePrefix = "arXiv",
    primaryClass = "nucl-th",
    doi = "10.1016/j.physletb.2021.136519",
    journal = "Phys. Lett. B",
    volume = "820",
    pages = "136519",
    year = "2021"
}

@article{Liu:2020ymh,
    author = "Liu, Yu-Chen and Mameda, Kazuya and Huang, Xu-Guang",
    title = "{Covariant Spin Kinetic Theory I: Collisionless Limit}",
    eprint = "2002.03753",
    archivePrefix = "arXiv",
    primaryClass = "hep-ph",
    doi = "10.1088/1674-1137/abae4d",
    journal = "Chin. Phys. C",
    volume = "44",
    number = "9",
    pages = "094101",
    year = "2020"
}

@article{Zubarev:1979,
    author = "Zubarev, D. N. and Prozorkevich, A. V. and Smolyanskii, S. A.",
    title = "{Derivation of nonlinear generalized equations of quantum relativistic hydrodynamics}",
    journal = "Theor. Math. Phys.",
    volume = "40",
    pages = "821--831",
    year = "1979",
    doi = "10.1007/BF01032069"
}

@article{Florkowski:2018fap,
    author = "Florkowski, Wojciech and Kumar, Avdhesh and Ryblewski, Radoslaw",
    title = "{Relativistic hydrodynamics for spin-polarized fluids}",
    eprint = "1811.04409",
    archivePrefix = "arXiv",
    primaryClass = "nucl-th",
    doi = "10.1016/j.ppnp.2019.07.001",
    journal = "Prog. Part. Nucl. Phys.",
    volume = "108",
    pages = "103709",
    year = "2019"
}

@article{Becattini:2022zvf,
    author = "Becattini, Francesco",
    title = "{Spin and polarization: a new direction in relativistic heavy ion physics}",
    eprint = "2204.01144",
    archivePrefix = "arXiv",
    primaryClass = "nucl-th",
    doi = "10.1088/1361-6633/ac97a9",
    journal = "Rept. Prog. Phys.",
    volume = "85",
    number = "12",
    pages = "122301",
    year = "2022"
}

@article{petz1993bogoliubov,
    author = "Petz, D. and Toth, G.",
    title = "{The Bogoliubov inner product in quantum statistics}",
    journal = "Lett. Math. Phys.",
    volume = "27",
    pages = "205--216",
    year = "1993",
    doi = "10.1007/BF00739578"
}

@book{petz2007quantum,
    author = "Petz, D{\'e}nes",
    title = "{Quantum Information Theory and Quantum Statistics}",
    series = "Theoretical and Mathematical Physics",
    publisher = "Springer",
    address = "Berlin, Heidelberg",
    year = "2008",
    doi = "10.1007/978-3-540-74636-2"
}

@article{Becattini:2025oyi,
    author = "Becattini, Francesco and Singh, Rajeev",
    title = "{On the local thermodynamic relations in relativistic spin hydrodynamics}",
    eprint = "2506.20681",
    archivePrefix = "arXiv",
    primaryClass = "nucl-th",
    journal = "Eur. Phys. J. C",
    volume = "85",
    number = "11",
    pages = "1338",
    year = "2025"
}

@article{Belinfante:1940,
    author = "Belinfante, F. J.",
    title = "{On the current and the density of the electric charge, the energy, the linear momentum and the angular momentum of arbitrary fields}",
    journal = "Physica",
    volume = "7",
    pages = "449--474",
    year = "1940",
    doi = "10.1016/S0031-8914(40)90091-X"
}

@article{Rosenfeld:1940,
    author = "Rosenfeld, L.",
    title = "{Sur le tenseur d'impulsion-\'energie}",
    journal = "Mem. Acad. Roy. Belg. Cl. Sci.",
    volume = "18",
    pages = "1--30",
    year = "1940"
}

@book{deGroot:1980,
    author = "de Groot, S. R. and van Leeuwen, W. A. and van Weert, Ch. G.",
    title = "{Relativistic Kinetic Theory: Principles and Applications}",
    publisher = "North-Holland",
    address = "Amsterdam",
    year = "1980"
}

@article{Hilgevoord:1963,
    author = "Hilgevoord, J. and Wouthuysen, S. A.",
    title = "{On the spin angular momentum of the Dirac particle}",
    journal = "Nucl. Phys.",
    volume = "40",
    pages = "1--12",
    year = "1963",
    doi = "10.1016/0029-5582(63)90246-X"
}

@article{Foldy:1950,
    author = "Foldy, L. L. and Wouthuysen, S. A.",
    title = "{On the Dirac theory of spin 1/2 particles and its non-relativistic limit}",
    journal = "Phys. Rev.",
    volume = "78",
    pages = "29--36",
    year = "1950",
    doi = "10.1103/PhysRev.78.29"
}

@article{Sheng:2022wsy,
    author = "Sheng, Xin-Li and Oliva, Lucia and Liang, Zuo-Tang and Wang, Qun and Wang, Xin-Nian",
    title = "{Spin alignment of vector mesons in heavy-ion collisions}",
    eprint = "2205.15689",
    archivePrefix = "arXiv",
    primaryClass = "nucl-th",
    doi = "10.1103/PhysRevLett.131.042304",
    journal = "Phys. Rev. Lett.",
    volume = "131",
    number = "4",
    pages = "042304",
    year = "2023"
}

@article{Sheng:2022ffb,
    author = "Sheng, Xin-Li and Oliva, Lucia and Liang, Zuo-Tang and Wang, Qun and Wang, Xin-Nian",
    title = "{Relativistic spin dynamics for vector mesons}",
    eprint = "2206.05868",
    archivePrefix = "arXiv",
    primaryClass = "hep-ph",
    reportNumber = "USTC-ICTS/PCFT-22-18",
    year = "2022"
}

@article{Xia:2020tyd,
    author = "Xia, Xiao-Liang and Li, Hui and Huang, Xu-Guang and Huang, Huan Zhong",
    title = "{Local spin alignment of vector mesons in relativistic heavy-ion collisions}",
    eprint = "2010.01474",
    archivePrefix = "arXiv",
    primaryClass = "nucl-th",
    doi = "10.1016/j.physletb.2021.136325",
    journal = "Phys. Lett. B",
    volume = "817",
    pages = "136325",
    year = "2021"
}

@article{Chen:2023hqi,
    author = "Chen, Jinhui and Liang, Zuo-Tang and Ma, Yu-Gang and Wang, Qun",
    title = "{Global spin alignment of vector mesons and strong force fields in heavy-ion collisions}",
    eprint = "2305.09114",
    archivePrefix = "arXiv",
    primaryClass = "nucl-th",
    doi = "10.1016/j.scib.2023.04.001",
    journal = "Sci. Bull.",
    volume = "68",
    pages = "874--877",
    year = "2023"
}

@article{Kumar:2023ghs,
    author = "Kumar, Avdhesh and Muller, Berndt and Yang, Di-Lun",
    title = "{Spin alignment of vector mesons by glasma fields}",
    eprint = "2304.04181",
    archivePrefix = "arXiv",
    primaryClass = "nucl-th",
    doi = "10.1103/PhysRevD.108.016020",
    journal = "Phys. Rev. D",
    volume = "108",
    number = "1",
    pages = "016020",
    year = "2023"
}

@article{Shen:2021jcg,
    author = "Shen, Diyu and Chen, Jinhui and Lin, Zi-Wei",
    title = "{The effect of hadronic scatterings on the measurement of vector meson spin alignments in heavy-ion collisions}",
    eprint = "2102.05266",
    archivePrefix = "arXiv",
    primaryClass = "nucl-ex",
    doi = "10.1088/1674-1137/abe763",
    journal = "Chin. Phys. C",
    volume = "45",
    number = "5",
    pages = "054002",
    year = "2021"
}

@article{Yang:2017sdk,
    author = "Yang, Yang-Guang and Fang, Ren-Hong and Wang, Qun and Wang, Xin-Nian",
    title = "{Quark coalescence model for polarized vector mesons and baryons}",
    eprint = "1711.06008",
    archivePrefix = "arXiv",
    primaryClass = "nucl-th",
    doi = "10.1103/PhysRevC.97.034917",
    journal = "Phys. Rev. C",
    volume = "97",
    number = "3",
    pages = "034917",
    year = "2018"
}

@article{Sheng:2019kmk,
    author = "Sheng, Xin-Li and Oliva, Lucia and Wang, Qun",
    title = "{What can we learn from the global spin alignment of $\phi$ mesons in heavy-ion collisions?}",
    eprint = "1910.13684",
    archivePrefix = "arXiv",
    primaryClass = "nucl-th",
    doi = "10.1103/PhysRevD.101.096005",
    journal = "Phys. Rev. D",
    volume = "101",
    number = "9",
    pages = "096005",
    year = "2020",
    note = "[Erratum: Phys.Rev.D 105, 099903 (2022)]"
}

@article{Liang:2004xn,
    author = "Liang, Zuo-Tang and Wang, Xin-Nian",
    title = "{Spin alignment of vector mesons in non-central A+A collisions}",
    eprint = "nucl-th/0411101",
    archivePrefix = "arXiv",
    reportNumber = "LBNL-56659",
    doi = "10.1016/j.physletb.2005.09.060",
    journal = "Phys. Lett. B",
    volume = "629",
    pages = "20--26",
    year = "2005"
}

@article{Becattini:2024aaa,
    author = "Becattini, Francesco and Buzzegoli, Matteo and Inghirami, Gabriele and Karpenko, Iurii and Palermo, Andrea",
    title = "{Local polarization and isothermal local equilibrium in relativistic heavy ion collisions}",
    eprint = "2103.14621",
    archivePrefix = "arXiv",
    primaryClass = "nucl-th",
    doi = "10.1103/PhysRevLett.127.272302",
    journal = "Phys. Rev. Lett.",
    volume = "127",
    number = "27",
    pages = "272302",
    year = "2021"
}

@article{Yang:2026weyl,
    author = "Yang, Shi-Zheng and Gao, Jian-Hua and Liang, Zuo-Tang and Prokhorov, Georgy Yu. and Pu, Shi and Teryaev, Oleg V. and Zakharov, Valentin I.",
    title = "{Weyl anomaly induced transport in hydrodynamics}",
    eprint = "2604.23849",
    archivePrefix = "arXiv",
    year = "2026"
}

@article{Jaynes:1957zza,
    author = "Jaynes, E. T.",
    title = "{Information Theory and Statistical Mechanics}",
    doi = "10.1103/PhysRev.106.620",
    journal = "Phys. Rev.",
    volume = "106",
    pages = "620--630",
    year = "1957"
}

@article{Jaynes:1957zz,
    author = "Jaynes, E. T.",
    title = "{Information Theory and Statistical Mechanics. II}",
    doi = "10.1103/PhysRev.108.171",
    journal = "Phys. Rev.",
    volume = "108",
    pages = "171--190",
    year = "1957"
}






\end{document}